\documentclass[12pt]{article}

\usepackage[reqno]{amsmath}
\usepackage[all]{xy}
\usepackage{amssymb}
\usepackage{textcomp}
\usepackage{amsmath}
\usepackage{color}

\newtheorem{thm}{Theorem}
\newtheorem{prop}{Proposition}

\newtheorem{lem}{Lemma}

\newtheorem{rem}{Remark}
\newtheorem{assumption}{Assumption}

\newcommand{\Z}{\mathbb{Z}}

\newcommand{{{\Ge}}}{\operatorname{G}}

\newcommand{\de}{\operatorname{d}}

\newcommand{{{\Ll}}}{\operatorname{L}}
\newcommand{{{\sgn}}}{\operatorname{sgn}}
\newcommand{{{\per}}}{\operatorname{per}}
\newcommand{{{\Rr}}}{\operatorname{R}}
\newcommand{{\Tr}}{\operatorname{Tr}}
\newcommand{\sumtwo}[2]{\sum_{\substack{#1 \\ #2}}}

\def\Re{{\operatorname{Re\,}}}

\newcommand{\M}{{\cal M}}

\newcommand{\C}{{\cal C}}

\begin{document}

\title{Geometric expansion of the log--partition function of the anisotropic Heisenberg model}
\author{Daniel Gandolfo$^{(1)}$, Suren Poghosyan$^{(2)}$, and Jean Ruiz$^{(3)}$}

\maketitle

  \parskip 5pt
%
%
%
%

\begin{abstract}

 We study the asymptotic expansion of the log-partition function of the anisotropic Heisenberg
 model in a bounded domain  as this domain is dilated to infinity. 
 Using the Ginibre's representation of the
 anisotropic Heisenberg model as a gas of interacting trajectories of a compound Poisson process
  we find all the non-decreasing terms of this expansion.
 They are given explicitly in terms of functional integrals.
 As the main technical tool we use the cluster expansion method.
\end{abstract}

\renewcommand{\thefootnote}{\arabic{footnote}}

\setcounter{footnote}{0}
\footnotetext[1]{
Centre de Physique Th\'{e}orique,
Aix--Marseille Univ, CNRS UMR 7332,
Univ Sud Toulon Var,
F-13288 Marseille Cedex 9, France.
\hfill\break
E-mail address: \texttt{gandolfo@cpt.univ-mrs.fr}
}
\footnotetext[2]{
Institute of Mathematics,
Armenian National Academy of Science,
Marshal Bagramian 24--B, Yerevan 375019, Armenia
\hfill\break
E-mail address: \texttt{suren.poghosyan@unicam.it}
}
\footnotetext[3]{
Centre de Physique Th\'{e}orique,
Aix--Marseille Univ, CNRS UMR 7332,
Univ Sud Toulon Var,
F-13288 Marseille Cedex 9, France.
\hfill\break
E-mail address: \texttt{ruiz@cpt.univ-mrs.fr}
}

\section{Introduction}

The problem of study of the large volume asymptotic behavior of the log-partition function for classical particle systems
in a bounded domain $\Lambda$
goes back to Van-Hove and Lee and Yang.
The  bulk term of the asymptotics was obtained by Van Hove in 1949 for the canonical ensemble \cite{VanH} and independently in 1952 by Yang and Lee \cite{LeeY} for the grand canonical ensemble. In the 1960's Ruelle and Fisher  gave the
mathematically rigorous proof of the existence of the
thermodynamic limit for classical particle systems ( see\cite{FishRu} and references therein).

The problem of finding the next terms of the asymptotic expansion was set up by Lebowitz in 1968 \cite{Leb}. Bounds for the difference between the log-partition function and the bulk term of its asymptotics can be found, for example in \cite{AGM} and \cite{Min-Sin}. A precise expression for the second term of the asymptotics in the case of lattice systems was found by Dobrushin in 1972 \cite{Dobr}.
Some generalization of Dobrushin's result is contained in \cite{FC}.

 A method which permits to find
 all the non-decreasing terms of the asymptotic expansion of the grand canonical log-partition function
 $\ln Z(\Lambda)$, when $\Lambda$ tends to the whole space by dilation,
was presented in \cite{P}. The method applies to classical continuous and discrete systems. This approach is based on the so called strong cluster estimates  of truncated correlation functions of the system
(see for instance \cite{DS} and references therein).
 The paper of Collet and Dunlop \cite{CD} extends the previous result and simplifies the proof for the case of classical continuous systems.

For quantum systems the first result   was presented by Macris, Martin and Pule in 1997 \cite{MMP}.
They considered a large volume asymptotics of Brownian integrals and derived the first three terms of the asymptotic expansion. For a special choice of the integrand this Brownian integral can be identified with the log-partition function $\ln Z_{\rm id}(\Lambda)$ of an ideal quantum gas
in Feynman-Kac  representation (so called loop gas).
It is worth to note that in contrast to the classical case 
$\ln Z_{\rm id}(\Lambda)$
has a non-trivial asymptotic expansion.

In \cite{PZ} 
a new approach is presented to the large volume asymptotic expansion of   $\ln Z(\Lambda)$ of an interacting loop gas
 in a bounded domain.  This approach, which can be applied to classical particle systems as well, is based on the abstract cluster expansion method \cite{PU} and uses estimates
of two-point truncated correlation functions only.

The same approach is used in the present paper to study the asymptotics  of the  log-partition function of a quantum
anisotropic Heisenberg model in a bounded domain.  We use Ginibre's functional integral representation based on a compound Poisson process \cite{G}.
This is a model of interacting closed trajectories (loops) in $\Z^d$ with random time intervals that are integer multiples
of $\beta$.  Each trajectory is constant except jumps, the possible values of jumps are
 vectors in $\Z^d$. The number of jumps in any finite time interval is finite with a
Poisson distribution.

Under general conditions on  transverse and longitudinal potentials of the Heisenberg model we establish the asymptotic behavior of
$\ln Z(\Lambda_R,z)$ as $ R\rightarrow \infty $ where   $\Lambda_R = \{R \cdot r\, |\,r\in\Lambda\}$,   $\Lambda \subset \Z^3$
is a  parallelepiped and $z$ is the fugacity (activity).
 The expansion is obtained up to order $o(1)$ and this constitutes the main result of the paper.
 All the non-decreasing terms of the expansion are proportional respectively to the volume, the area of the boundary
 the total length of the edges and the number of vertices of $\Lambda_R$.
 The coefficients which  depend on the potentials are given explicitly as functional integrals and are analytic in $z$ in a neighborhood of the origin.
This expansion assumes  low fugacity and invariance of the potentials with respect to the automorphism group of the lattice $\Z^3$.

We consider the case $d=3$ only for simplicity, all the arguments can be easily adopted for the case of arbitrary $d\geq 1$.
An important technical tool for the analysis of the asymptotics of $\ln Z(\Lambda_R,z)$ is Proposition 1 which describes a
decay property of the two-point truncated correlation function for the corresponding system of interacting loops.

The paper is organized as follows. Section II considers the model and presents the main result of the paper.
In Section III we describe the corresponding loop model
as an interacting gas of closed trajectories of a compound Poisson process on $\Z^d$.
Section IV presents cluster expansions and decay of correlations in frame of our loop model. The proof of the main result is given
in Section V.

\section{Main result}

\medskip

 It was noted by Matsubara and Matsuda \cite{MM} that one may think of the anisotropic Heisenberg model as a quantum-mechanical lattice gas.
Following Ginibre \cite{G} we take this point of view and introduce at each site $r$ of the $d$-dimensional cubic lattice
$r\in{\Z}^d$ Boson field operators  $a_r, a_r^+$ which  annihilates and  creates a particle at $r$-th lattice point.
They satisfy the following commutation relations:
$$[a_r,a_s]=0,\,\,[a^+_r,a^+_s]=0,\,\,[a_r,a^+_s] =\delta_{r,s}, $$
where $[A,B]=AB - BA$ is the commutator of two operators.

 We consider a system confined in a parallelepiped $\Lambda\subset \Z^d$ and describe it in the grand canonical formalism.
The states of the  system  form a vector space $\mathcal{H}_\Lambda$ with a normalized basis, labeled by the subsets of $\Lambda$. (see for details  \cite{GMR} or \cite{Rob})
The  Hamiltonian  is given by
\begin{eqnarray}
H_\Lambda=-\frac{1}{4}\sumtwo{r, s\in\Lambda}{r\neq s}(a_r^+-a_s^+)(a_r-a_s)\pi(r-s)\nonumber\\
+\frac{1}{2}\sumtwo{r, s\in\Lambda}{r\neq s}a_r^+a_s^+a_ra_s \psi(r-s)
-\mu \sum_{r}a_r^+a_r.
\end{eqnarray}
Here $\pi$ and $\psi$ are real valued functions defined on $\Z^d$.
They are called the transverse and longitudinal potentials respectively, $\mu$ is the chemical potential.

The main object of our interest is the  grand partition function  $Z(\Lambda_R, z)=\Tr e^{-\beta H_{\Lambda_R}}$.
To study the asymptotic behavior of
  $\ln Z(\Lambda_R, z)$ as $R\rightarrow\infty$ we use a functional integral representation of $Z(\Lambda_R, z)$
  developed by Ginibre \cite{G}.

 We assume that $\pi$ is symmetric: $\pi(r) = \pi(-r)$ for all $r\in \Z^d$
with
\begin{equation}
\sum_{r\in\Z^d\setminus \{0\}} |\pi(r)|=2M<\infty
\end{equation}
and let
\begin{equation}
\sum_{r\in\Z^d\setminus \{0\}} \pi(r)=2M_0.
\end{equation}
We assume also that the longitudinal potential  has a hard core
$\psi(0)=+\infty,$
is symmetric and satisfies
$\sum_{r\in\Z^d\setminus \{0\}} |\psi(r)|<\infty.$

 We will deal with complex $\pi$ and $\psi$ as well. The set of complex potentials $\vartheta$ form a Banach space with the norm
$ ||\vartheta||=\sum_{r\in\Z^d\setminus \{0\}}|\vartheta(r)|$.

 Let
 \begin{equation}
\Lambda = \{r=(r_1,r_2,r_3)\in \Z^3\,\,\mid\,\,0\leq r_i\leq a_i,\,\,a_i>0,\,\,i=1,2,3 \}.
\end{equation}

We denote by $\Lambda^{(f)},\,\,\Lambda^{(e)}$ and $\Lambda^{(v)}$  respectively the set of all faces, edges and vertices of the parallelepiped $\Lambda$.
The main result of this paper is
\begin{thm}
Let the transverse and longitudinal potentials $\pi$ and $\psi$ satisfy respectively the conditions
\begin{equation}
2M_l = \sum_{r\in\Z^3\setminus \{0\}} |\pi(r)| (1+|r|)^l  <\infty,\,\,\,\,l>3
\label{Mdelta}
\end{equation}
and
\begin{equation}
||\psi_l||= \sum_{r\in\Z^3\setminus \{0\}} |\psi_l(r)|  <\infty,\,\,\,\,l>3
\label{psidelta}
\end{equation}
with $\psi_l(r)=\psi(r)(1+|r|)^l$. Then for all $z$ such that
\begin{equation}
p_l(z)=C(\beta,l)\sum_{j=1}^\infty j [|z|  e^{\beta(M_l e+\Re M_0+ 2||\psi||) +1}]^j <1
\label{p-forAss4}
\end{equation}
with
\begin{equation}
 C(\beta,l)= (1+\beta M_l e)(2+2\beta ||\psi_l||  +\beta M_l e )
 \label{Const.inAss4}
\end{equation}
the following asymptotic expansion holds true
\begin{eqnarray}
\ln Z(\Lambda_R, z) = R^3 |\Lambda|A_0 (z) + R^2\sum_{\lambda\in\Lambda^{(f)} } |\lambda| A_1(\lambda, z)
+R \sum_{\lambda\in\Lambda^{(e)}} |\lambda|A_2(\lambda,z)\nonumber\\
 + \sum_{\lambda\in\Lambda^{(v)}} A_3(\lambda,z) + o(1),\,\,\,\,\,R\rightarrow\infty.
\end{eqnarray}
Here $|\cdot|$ denotes the number of elements in a finite set, the coefficients $A_j(\Lambda,z)$ are given explicitly below
 by Eqs. (\ref{A0}), (\ref{A1}), (\ref{A12}) and (\ref{I123Final}). They all are analytic in $z$ in the circle (\ref{p-forAss4}). Note that $\beta^{-1}A_0(z)$ is the grand canonical pressure.

If the potentials $\pi$ and $\psi$ are in addition invariant with respect to the automorphism group of the lattice $\Z^3$
then the coefficients  $A_j(\lambda, z),\,j=1,2,3$   take a simpler form
\begin{eqnarray}
\ln Z(\Lambda_R, z) = R^3 |\Lambda|A_0 (z) + R^2\sum_{\lambda\in\Lambda^{(f)} } |\lambda| A_1( z)
+R \sum_{\lambda\in\Lambda^{(e)}} |\lambda|A_2(z)\nonumber\\
 + 8 A_3(z) + o(1),\,\,\,\,\,R\rightarrow\infty
\end{eqnarray}
with $A_j(z)$ given by Eqs. (\ref{A^b}), (\ref{A^e}) and (\ref{A^v}).

Note that $\sum_{\lambda\in\Lambda^{(f)} } = |\partial\Lambda|$ is the total area of the boundary of $\Lambda$,
$\sum_{\lambda\in\Lambda^{(e)}}$ is the total length of the edges of the parallelepiped $\Lambda$.
\end{thm}

We formulate and prove the main result for the case $d=3$ only for simplicity. The arguments used for the proof can be directly applied in higher dimensions. Having this in mind we present the loop model, the cluster expansions and the decay of correlations in general case $d\geq 1$.

\section{Loop model}

As it was shown by Ginibre \cite{G} the statistical operator $e^{-\beta H_\Lambda}$ of the anisotropic Heisenberg model can be represented as a functional integral in the space of paths of a \emph{compound Poisson process} with intensity $M$ given by (2).

The compound Poisson process on ${\Z}^d$ with intensity $M > 0$ and jump size
distribution $p$ is a stochastic process $X(t)$ defined as
\begin{equation}
X(t)=\sum_{i=1}^{N_t}Y_i,
\end{equation}
where jump sizes $Y_i$ are independent identically distributed ${\Z}^d$-valued random variables with common   distribution
$p=(p_k)_{k\in{\Z}^d};\,\sum_{k\in{\Z}^d}p_k =1$
and $N_t$ is a Poisson process with intensity $M$, independent from $(Y_i)_{i\geq1}$ (see for instance \cite{Appl}).
Hence $X(t)$  is the stochastic
process which starts at 0, stays there for an exponential holding
time with mean value $M^{-1}$, then it jumps by a vector $k\in{\Z}^d$
with probability $p(k)$, stays at $k$ for another, independent holding
time with mean $M^{-1}$, jumps again,
etc. The number of jumps in any finite time interval is finite with Poisson distribution of intensity $M$.

\medskip

The jump distribution $p$ is defined with the help of the transverse potential $\pi$
of the Heisenberg model.
Following Ginibre \cite{G}
we first  consider the case where $\pi$ is non-positive, $\pi\leq 0$. As it is mentioned by Ginibre \cite{G} this condition means that the
 ``mass of the particle" is positive.
In this case the jump distribution
 \begin{equation}
 p(r) = \frac{-\pi(r)}{2M}.
 \end{equation}

Let
$${\mathcal X}^{t}_{r,s}=\{X:[0,t]\rightarrow \Z^d\,\,\mid\,\, X(0)=r, X(t) =s\},\,\,t>0,\,\,r,s \in \Z^d$$
 be the set of right-continuous with left limits  piecewise constant functions (trajectories) defined on the interval $[0,t]$
with values in $\Z^d$. We call $t$ the length of the trajectory $X\in {\mathcal X}^{t}_{r,s}$.
Trajectories from $\mathcal X^{t}_{r,s}$ can be parametrized by giving the number $n$ of jumps, the jumping times $t_1\leq t_2\leq\cdots\leq t_n$ and the successive jumps $r_1,r_2,\cdots,r_n$ such that $r+r_1+r_2+\cdots +r_n=s$.
On  $\mathcal X^{t}_{r,s}$  we consider a path measure
$P^t_{r,s}$ given by
\begin{eqnarray}
\int_{\mathcal X^{t}_{r,s}} P^t_{r,s} (\de X) h(X)=\sum_{n=0}^\infty\sumtwo{r_1,r_2,\cdots,r_n\in \Z^d}{r+r_1+r_2+\cdots +r_n=s}
\prod_{i=1}^n \frac{-\pi(r_i)}{2M}\nonumber\\
\cdot M^{-1}\int_{[0,t]^n}\de t_1\cdots \de t_n \prod_{i=0}^{n}g_M(t_{i+1}-t_i) h(X)
\end{eqnarray}
 where we used the convention $t_0=0,\,t_{n+1}=t$.  Here $h$ is any non-negative function on $\mathcal X^{t}_{r,s}$ and
\begin{equation}
g_M(s)= \begin{cases}M e^{-Ms},\,s\geq0\\0,\,s<0 \end{cases}.
\end{equation}
is the density of the exponential distribution with mean $M^{-1}$.
Note that $\sum_{s\in \Z^d} P^t_{r,s}(\mathcal X^t_{r,s})   = 1.$

We pass to the case of complex measure $P^t_{r,s}$.  Now $\pi$ is not necessarily negative and can be also complex.
To such potentials we associate a complex measure $P^t_{r,s}$  as follows. Let $P^t_{r,s,+}$ be the non-negative measure  defined by means of
$|\pi|$ as was described above.
We define the measure $P^t_{r,s}$ on the same space of paths ${\mathcal{X}^t_{r,s}}$ so that it is absolutely continuous with respect to the measure
$P^t_{r,s,+}$ and its Radon-Nikodym derivative $f_t (X)$ is defined almost everywhere (with respect to $P^t_{r,s,+}$) by
\begin{equation}
f_t(X) =\exp[t(M+M_0)] \prod_{i=1}^{n}\frac{-\pi(r_i)}{|\pi(r_i)|}
\end{equation}
if the path $X$ has $n$ jumps of magnitude  $r_1,\cdots,r_n$ in time interval $(0,t)$.


If the function $h(X),\,X\in\mathcal X^{t}_{r,s},$ depends only on the number of jumps $n$ and the magnitudes $r_1,r_2,\cdots,r_n$
then 
\begin{align}
\int_{\mathcal X^{t}_{r,s}} P^t_{r,s} (\de X) h(X) = e^{-Mt}\sum_{n=0}^\infty \frac{(Mt)^n}{n!}
\sumtwo{r_1,r_2,\cdots,r_n\in \Z^d}{r+r_1+r_2+\cdots +r_n=s}
\prod_{i=1}^n \frac{-\pi(r_i)}{2M}\nonumber\\
\cdot h(r_1,r_2,\cdots,r_n).
\label{onlyjumps}
\end{align}

It is worth to note that $|f_t(\omega)| =\exp[t(M+ Re M_0)]$ and $f_t\equiv 1$ if $\pi$ is non-positive.
Therefore the following equation is valid
\begin{equation}
| P^t_{r,s}| =\exp[t(M+ Re M_0)]  P^t_{r,s,+}
\end{equation}
where $|P|$ denotes the total variation of the complex measure $P$.

Let $\beta>0$ be fixed and let
\begin{equation}
\mathcal{X}^{j\beta} = \sum_{r\in\Z^d}{\mathcal X}^{j\beta}_{r,r},\,\,\,j=1,2,\cdots
\end{equation}
be the set of all closed trajectories of length $j\beta$.
We set $\mathcal{X} = \sum_{j = 0}^{\infty} \mathcal{X}^{j\beta}$
and use $\mathcal{X} ( \Lambda )$ to denote respectively the set of all
closed trajectories of length multiple to the fixed parameter $\beta>0$ in ${\Z}^d$ or lying in a
subset $\Lambda \subset {\Z}^d$. The elements of $\mathcal{X}$ we will call composite trajectories (composite loops).
The elements of $\mathcal{X}_{r,s}^{\beta}$ we call elementary trajectories. We will say that  $x\in \mathcal{X}_{r,s}^{\beta}$ is an elementary constituent of a composite trajectory $X\in \mathcal{X}^{j\beta},\,\,j>1$ and write $x\in X$ if $x(s)= X(s+k\beta)$ for some $0\leq k\leq j-1$.

Based on the compound Poisson process Ginibre \cite{G} gave a functional integral representation  of the partition function $Z(\Lambda,z)$ of the anisotropic Heisenberg model confined in a bounded region $\Lambda\subset\Z^d$. This representation is given in terms of composite loops
and has the form:
\begin{eqnarray}
Z(\Lambda,z) =\sum_{n\geq0} \frac{1}{n!}\int_{\mathcal{X} ( \Lambda )} \bar{\mu}_z(\de X_1)\cdots \int_{\mathcal{X} ( \Lambda )} \bar{\mu}_z(\de X_n)
 e^{-U(X_1,\cdots,X_n)}.
\end{eqnarray}
Here
\begin{equation}
U(X_1,\cdots,X_n)=\sum_{i<j}^n u(X_i,X_j) + \sum_{i=1}^n v(X_i)
\end{equation}
with
\begin{equation}
v(X)=\frac{1}{2}\sum_{x,\bar{x}\in X,x\neq\bar{x}}\int_0^\beta \psi(x(t)-\bar{x}(t))\de t
\end{equation}
and
\begin{equation}
u(X,Y)=\sum_{x\in X}\sum_{y\in Y}\int_0^\beta \psi(x(t)-y(t))\de t
\end{equation}
where $\psi$ is the longitudinal potential.
The measure $\bar{\mu}_z$ is given by
\begin{eqnarray*}
\int_{\mathcal{X} ( \Lambda )}\bar{\mu}_z(\de X) h(X)
=\sum_{r\in\Lambda}\sum_{j\geq1}\frac{z^{j}}{j}\int P^{j\beta}_{r,r}(\de X)1_{\mathcal{X} ( \Lambda )}(X) h(X),
\end{eqnarray*}
where $h$ is any non-negative function on  $\mathcal{X}(\Lambda)$ and $1_{E}$ is the indicator function of the set $E$.

\section{Cluster expansion and decay of correlations}

To prove Theorem 1 we undertake the following strategy. As a first step we establish with the help of Theorem 2.1 from \cite{PU} the cluster representation of the partition function in terms of the Ursell function. This allows to write the log-partition function as an absolute convergent integral of the Ursell function over the space of finite configurations of loops in $\Lambda_R$. Then we get the bulk term by separating a loop which stays in $\Lambda_R$ and releasing all other constraints. To get the further terms of the expansion we need to study the decay of correlations of our model of interacting loops. This is done with the help of Theorem 2.3 from \cite{PU} and its further modification.

To apply the abstract cluster expansion method \cite{PU} we rewrite the partition function  in the following way
\begin{equation}
Z(\Lambda,z) =\sum_{n\geq0} \frac{1}{n!}\int_{\mathcal{X} ( \Lambda )} \mu_z(\de X_1)\cdots \int_{\mathcal{X} ( \Lambda )} \mu_z(\de X_n)
\exp\left\{-\sum_{i<j}^n u(X_i,X_j)\right\}
\end{equation}
where the measure $\mu_z$ is given by
\begin{eqnarray}
\mu_z(\de X) = e^{-v(X)} \bar{\mu}_z(\de X).
\end{eqnarray}
Note that the total variation $|\mu_z|$  of the measure $\mu_z$ is a locally finite measure for $z$ small enough. More precisely
for all finite $\Lambda$
\begin{equation}
|\mu_z|(\mathcal X (\Lambda))< |\Lambda|  \sum_{j\geq 1}
\frac{[z e^{\beta(M +\Re M_0 + ||\psi||)}]^j}{j} <\infty
\label{estmodmuz}
\end{equation}
provided $z e^{\beta(M +\Re M_0 + ||\psi||)}<1$.

\begin{rem}
We note that the self energy $v(X)= +\infty$ whenever two elements of the collection of elementary paths that constitute the composite loop $X$ have overlapping hard cores (are on the same lattice site for some time interval).
The set of such loops has $\mu_z$ measure zero.
Therefore from now on we will consider only the subspace of those composite  loops from $\mathcal{X}$ any two elementary constituents of which  have non-overlapping hard cores. We will call such loops admissible and use the same notations
$\mathcal{X}^{j\beta}_{rr}$  or $\mathcal{X}(\Lambda)$ for corresponding subspaces of admissible loops.
\end{rem}

Theorem 2.1 from \cite{PU} holds true under two assumptions on the interaction $u$. The first one assumes the stability on the potential $u$.
\begin{assumption}
There exists a nonnegative function $b$ on $\mathcal{X}( \Lambda )$ such that, for all $n$ and all admissible
$X_1,\dots,X_n \in \mathcal{X} ( \Lambda )$,
\[
\sum_{1\leq i<j \leq n} \Re u(X_i,X_j) \geq \sum_{i=1}^n b(X_i).
\]
\end{assumption}
The next condition is related to the strength of the interaction $u$. Let $\zeta(X,Y)= e^{-u(X,Y)}-1$ be the Mayer function.
\begin{assumption}
There exists a nonnegative function $a$ on $\mathcal{X}( \Lambda )$ such that for all admissible  $X\in\mathcal{X}( \Lambda )$,
\[
\int_{\mathcal{X}( \Lambda )}|\mu_z|(\de Y) \, |\zeta(X,Y)| e^{a(Y)+2b(Y)} \leq a(X).
\]
\end{assumption}

 Theorem 2.1 from \cite{PU} states that if Assumptions 1 and 2 are valid and in addition
$$\int_{\mathcal X (\Lambda)} |\mu_z|(\de X)| e^{a(X)+2b(X)} < \infty$$
then
\begin{equation}
 Z(\Lambda,z) = \exp \left\{ \sum_{n\geq1}\frac{1}{n!} \int\mu_z(\de X_1) \dots \mu_z(\de X_n) \, \varphi(X_1,\dots,X_n)\right\} .
\label{lnZZ}
\end{equation}
 The term in the exponential converges absolutely.
Here $\varphi$ is the Ursell function given by
\begin{equation}
\varphi(X_1,\dots,X_n) = \begin{cases} 1 & \text{if } n=1, \\  \sum_{G \in \C_n} \prod_{\{i,j\} \in G} \zeta(X_i,X_j) & \text{if } n\geq2 \end{cases}
\end{equation}
where $\C_n$ is the set of all connected graphs with $n$ vertices and the product is over edges of $G$.

We will use for convenience slightly stronger condition than Assumption 2.
\begin{assumption}
There exists a nonnegative function $a$ on $\mathcal{X}( \Lambda )$ and a number $p,\,\,0<p<1$,
such that for all admissible  $X\in\mathcal{X}( \Lambda )$,
\[
\int_{\mathcal{X}( \Lambda )}|\mu_z|(\de Y) \, |\zeta(X,Y)| e^{a(Y)+2b(Y)}a(Y) \leq p a(X).
\]
\end{assumption}

We show in the Appendix A2 that Assumption 1 holds with
\begin{equation}
b(X)= \frac{1}{2} \Re  v(X) +\frac{1}{2}\beta ||\psi|| \mid X \mid
\label{functionb}
\end{equation}
where $|X|$  denotes the length of the composite loop $X,\,\, |X|=j$ if $X\in\mathcal{X}^{j\beta}$.
In the Appendix A3 we prove that Assumption 3 (hence also Assumption 2) holds true for sufficiently small $z$
with the following choice of the function $a$:
\begin{equation}
a(X)= |X| + N(X)
\label{functiona}
\end{equation}
where $N(X)$ is the number of jumps of  $X$.

Let  $\mathcal M = \sum_{n=0}^\infty {\mathcal X}^{\otimes n}$
be the space  of finite sequences of loops where by definition $ {\mathcal X}^{\otimes0 } =\{\emptyset\} $ is a singleton.
The measure $\mu_z$ generates in a canonical way  a  measure $ W_{\mu_z}$ on  $\mathcal M$ given by
\begin{equation}
 W_{\mu_z} = \sum_{n=0}^\infty \frac{1}{n!}{\mu_z}^{\otimes n}
\end{equation}
where for each $n$ the product measure ${\mu_z}^{\otimes n}$ is supported by $ {\mathcal X}^{\otimes n}$ with ${\mu_z}^{\otimes 0}(\{\emptyset\})\\
=1$.

For  $\Lambda\subset \Z^d$ we denote by $W_{\mu_z,\Lambda}$  the restriction of the measure $W_{\mu_z}$ on the space $\M(\Lambda)$
of finite sequences of loops in $\Lambda$.  It follows from \ref{estmodmuz} that $W_{|\mu_z|,\Lambda}$
is a finite measure for all bounded $\Lambda$ if $z< e^{-\beta(M+\Re M_0 + ||\psi||)}$.  We note also that the total variation of the measure $W_{\mu_z}$ is $W_{|\mu_z|}$.

For any function $F(\omega_1; \omega_2),\,\,\,\omega_1, \omega_2 \in\M $ which is symmetric in both variables $\omega_1$ and
$\omega_2$ the following formula
\begin{equation}
 \int_{\M}W_{\mu_z} \de (\omega) \sum_{\omega_1\subset\omega} F(\omega_1;\omega\setminus\omega_1)
 = \int_{\M}W_{\mu_z} \de (\omega_1)  \int_{\M}W_{\mu_z} \de (\omega_2)  F(\omega_1;\omega_2)
\label{MM}
\end{equation}
holds true if either the functions $F$
is non-negative or at least one side is absolutely convergent. (See for example \cite{Rue}, section 4.4.)
Here the sum in the left side runs over all subsequences $\omega_1$ of $\omega$ and $\omega\setminus\omega_1$ is a subsequence
which is obtained from $\omega $ by deleting the elements of $\omega_1$.

To derive the asymptotic expansion we need certain decay of correlations of the system which is given in terms of bounds for
 the two-point truncated correlation functions $\sigma$. When the cluster expansion converges they are given by
 \begin{equation}
 \sigma(X,Y) = \int_{\M} W_{\mu_z}(\de \omega) \varphi(X,Y,\omega),\,\,\,X,Y \in {\mathcal X}.
 \end{equation}
 See \cite{PU}, Theorem 2.2.

 We will use the following notations
 \begin{equation}
 |\sigma|(X,Y) = \int_{\M} W_{|\mu_z|}(\de \omega) |\varphi|(X,Y,\omega)
 \label{nsigmaxy}
 \end{equation}
 with
 \begin{equation}
 |\varphi|(\omega) = \sum_{G \in \C_{|\omega|}} \prod_{\{i,j\} \in G} |\zeta(x_i,x_j) |.
 \end{equation}


We need one more assumption related to the decay of the interaction $u$.
\begin{assumption}
There exists a nonnegative function $a$ on $\mathcal{X}$ and a number $p,\,\,0<p<1$  such that
\[
\int_{{\mathcal X}^c(B_0(R+r))}\de|\mu_z|(Y) \, |\zeta(X,Y)| e^{a(Y) + 2b(Y)} a(Y) \leq p a(X) (1+r)^{-l},\,\,\,l>3
\]
for any $r>0$ and all admissible  $X \in {\mathcal X}(B_0(R))$.
\end{assumption}

The proof of Assumptions  4 is given in the Appendix A4.

\section{Proof of the Theorem}

We assume that the orientation of the  faces of the parallelepiped $\Lambda$ are given by the inward drawn unit normals $n_1, \cdots, n_6$.  Let
$\lambda=\lambda_l \in  \Lambda^{(f)}$ be the  face of $\Lambda$ which is defined by the normal $n_l$.
An edge of $\lambda\in\Lambda^{(e)}$ defined by a pair of adjacent faces $\lambda_{l_1},\lambda_{l_2} \in  \Lambda^{(f)}$ is denoted by
$\lambda_{l_1,l_2}$ so that $\lambda_{l_1,l_2} = \lambda_{l_1}\cap\lambda_{l_2}$.
In the same way by $\lambda_{l_1, l_2, l_3}$ we denote the vertex of $\Lambda$ where three faces $\lambda_{l_1}$,
$\lambda_{l_2}$ and $\lambda_{l_3}$ are meeting:  $\lambda_{l_1, l_2, l_3} =\bigcap_{i=1,2,3} \lambda_{l_i}$.

Let $(e_1,e_2,e_3)$ be the usual orthonormal basis in $\mathbb{R}^3$. We define the unit normals by $n_i=e_i,\,n_{i+3} = -e_i + a_i,\,\,i=1,2,3$
and we denote by $N_l$ the half spaces
\begin{equation}
N_l= \{r\in\Z^3\, |\, \langle r,n_l\rangle\geq 0 \},\,\,l=1,\cdots,6\}.
\end{equation}

In terms of the measure $W_{\mu_z}$ we can rewrite (\ref{lnZZ}) as
 \begin{equation}
\ln Z(\Lambda, z) =  \int_{\M} W_{\mu_z}(\de \omega) \varphi(\omega) 1_{\M(\Lambda)} (\omega)
\end{equation}
where
\begin{equation}
1_{\M(\Lambda)} (\omega)=\prod_{X\in\omega}1_{\mathcal{X}(\Lambda)}(X).
\end{equation}
We use the Heavyside function
\begin{equation}
H(s)= \begin{cases} 1,\,s\geq0\\0,\,s<0 \end{cases}
\end{equation}
to represent the indicator function as
\begin{equation}
1_{\M(\Lambda)} (\omega)=\prod_{i=1}^6 H(\inf\langle\omega,n_i\rangle)=\prod_{i=1}^6 [1-H(-\inf\langle\omega,n_i\rangle -1])
\label{indfunct}
\end{equation}
where
\begin{equation}
\inf\langle\omega,n_i\rangle=\min_{X\in\omega}\inf_{t\in[0, |X|\beta]}\langle X(t),n_i\rangle.
\end{equation}

With the help of the formula \ref{MM}   we separate a loop from a configuration $\omega$ and get the
representation
 \begin{multline}
\ln Z(\Lambda_R, z) = \int_{{\mathcal X}} \mu_z(\de X) 1_{\mathcal X(\Lambda_R)}(X) \int_{\M} W_{\mu_z}(\de \omega)
 \frac{\varphi(X,\omega)}{|\omega| +1} 1_{\M(\Lambda_R)} (\omega)\\
 =\sum_{r\in\Lambda_R} \sum_{j\geq1} \frac{z^j}{j}\int_{{\mathcal X}^{j\beta}_{00}} P^{j\beta}_{00} (\de X^0)e^{-v(X^0)}
 \int_{\M} W_{\mu_z}(\de \omega)\frac{\varphi(X^0 +r,\omega)}{|\omega| +1}\\
 \cdot 1_{\M(\Lambda_R)} (X^0+r,\omega).
 \label{lnZ}
\end{multline}

From (\ref{indfunct}) it follows that
\begin{align}
1_{\M(\Lambda_R)} (X,\omega) 
= 1 &-\sum_{l=1}^6
H(-\inf\langle(X,\omega),n_l\rangle -1)
\nonumber \\
&
+ \sideset{}{^{(\text{adj})}} \sum_{1\leq l_1<l_2\leq 6}
\prod_{i=1,2}H(-\inf\langle(X,\omega),n_{l_i}\rangle -1)
\nonumber \\
&-  \sideset{}{^{(\text{adj})}}\sum_{1\leq l_1<l_2<l_3\leq 6}
\prod_{i=1,2,3}H(-\inf\langle(X,\omega),n_{l_i}\rangle -1)
\nonumber \\
&
+\sum_{L^*\subset \{1,2,...,6\}}
(-1)^{|L^*|}
\prod_{l\in L^*}H(-\inf\langle(X,\omega),n_{l}\rangle -1).
\label{indfunct2}
\end{align}
Here the sums $\sideset{}{^{(\text{adj})}} \sum$ means that the faces $\lambda_{l_1}$ and  $\lambda_{l_2}$
(resp.\  $\lambda_{l_1},\,\lambda_{l_2}$ and $\lambda_{l_3}$) are adjacent and the last sum is restricted
to subsets $L^*$ for which at least two faces $\lambda_{l_1}$, $\lambda_{l_2},\, l_1,l_2 \in L^*$ are parallel.

Combining (\ref{lnZ}) and (\ref{indfunct2}) we can write
\begin{align}
\ln Z(\Lambda_R, z)&= I_0(R,z) + \sum_{l=1}^6 I_l(R,z) + \sideset{}{^{(\text{adj})}}\sum_{1\leq l_1<l_2\leq 6}I_{l_1,l_2}(R,z)\nonumber\\
 &+
 \sideset{}{^{(\text{adj})}}\sum_{1\leq l_1<l_2<l_3\leq 6}I_{l_1,l_2,l_3}(R,z) + Q(R,z)
 \label{expansion}
\end{align}
where
\begin{align}
I_0(R,z)=
\sum_{r\in\Lambda_R} \sum_{j\geq1} \frac{z^j}{j}\int_{{\mathcal X}^{j\beta}_{00}} P^{j\beta}_{00} (\de X^0)e^{-v(X^0)}
 \int_{\M} W_{\mu_z}(\de \omega)\frac{\varphi(X^0 +r,\omega)}{|\omega| +1},
 \label{I0}
\end{align}

\begin{align}
I_l(R,z)= &-
\sum_{r\in\Lambda_R} \sum_{j\geq1} \frac{z^j}{j}\int_{{\mathcal X}^{j\beta}_{00}} P^{j\beta}_{00} (\de X^0)e^{-v(X^0)}
 \int_{\M} W_{\mu_z}(\de \omega)\frac{\varphi(X^0 +r,\omega)}{|\omega| +1} \nonumber\\
& \cdot H(-\inf\langle(X^0+r,\omega),n_l\rangle -1) ,
\label{I1}
\end{align}

\begin{align}
I_{l_1,l_2}(R,z)= &
\sum_{r\in\Lambda_R} \sum_{j\geq1} \frac{z^j}{j}\int_{{\mathcal X}^{j\beta}_{00}} P^{j\beta}_{00} (\de X^0)e^{-v(X^0)}
 \int_{\M} W_{\mu_z}(\de \omega)\frac{\varphi(X^0 +r,\omega)}{|\omega| +1} \nonumber\\
&\cdot
\prod_{i=1,2}H(-\inf\langle(X,\omega),n_{l_i}\rangle -1)  ,
\label{I12}
\end{align}

\begin{align}
I_{l_1,l_2,l_3}(R,z)= &-
\sum_{r\in\Lambda_R} \sum_{j\geq1} \frac{z^j}{j}\int_{{\mathcal X}^{j\beta}_{00}} P^{j\beta}_{00} (\de X^0)e^{-v(X^0)}
 \int_{\M} W_{\mu_z}(\de \omega)\frac{\varphi(X^0 +r,\omega)}{|\omega| +1}\nonumber\\
&\cdot
\prod_{i=1,2,3}H(-\inf\langle(X,\omega),n_{l_i}\rangle -1).
\label{I123}
\end{align}
Finally
\begin{equation}
Q(R,z) =\sum_{L^*\subset \{1,2,...,6\}} I_{L^*}
\label{QRz}
\end{equation}

with
\begin{align}
I_{L^*}
&= (-1)^{|L^*|}  \sum_{r\in\Lambda_R} \sum_{j\geq1} \frac{z^j}{j}\int_{{\mathcal X}^{j\beta}_{00}} P^{j\beta}_{00} (\de X^0)e^{-v(X^0)}
 \int_{\M} W_{\mu_z}(\de \omega)\nonumber\\
&\cdot\frac{\varphi(X^0 +r,\omega)}{|\omega| +1} \prod_{l\in L^*} H(-\inf\langle(X^0+r,\omega),n_l\rangle -1).
\label{Itrash}
\end{align}

\subsection{The bulk term}

 By the translation invariance of the measure $W_{\mu_z}$ and the Ursell function
the bulk (volume) term of the geometric expansion is
\begin{align}
I_0(R,z)=
&\sum_{r\in\Lambda_R} \sum_{j\geq1} \frac{z^j}{j}\int_{{\mathcal X}^{j\beta}_{00}} P^{j\beta}_{00} (\de X^0)e^{-v(X^0)}
 \int_{\M} W_{\mu_z}(\de \omega)\frac{\varphi(X^0,\omega)}{|\omega| +1}\nonumber\\
 &=R^3 |\Lambda| A_0(z),
 \label{bulkterm}
\end{align}
where
\begin{align}
A_0(z) =
\sum_{j\geq1} \frac{z^j}{j}\int_{{\mathcal X}^{j\beta}_{00}} P^{j\beta}_{00} (\de X^0)e^{-v(X^0)}
 \int_{\M} W_{\mu_z}(\de \omega)\frac{\varphi(X^0,\omega)}{|\omega| +1}.
 \label{A0}
\end{align}
It is worth to note that the pressure $p_0(z) = p_0(z, \pi, \psi) = \beta^{-1} A_0(z)$ with $A_0(z)$ given by (\ref{A0}).
According to Theorem 2.1 from \cite{PU} for all admissible $X\in \mathcal X$,
\begin{equation}
\int_{\M} W_{|\mu_z|}(\de \omega)
\frac{|\varphi(X,\omega)|}{|\omega| +1}\leq e^{a(X) +2b(X)}.
\label{intdomega}
\end{equation}
Hence with the help of the equality
\begin{equation}
 e^{-\Re v(Y^0)} e^{a(Y^0)+2b(Y^0)} =e^{\beta ||\psi|| |Y^0| + |Y^0| + N(Y^0)}
 \label{Rev-a-b-norm}
 \end{equation}
 we have that
 \begin{align}
|A_0(z)| \leq
\sum_{j\geq1}\frac{1}{j} [z e^{\beta(M+\Re M_0 + ||\psi||) + 1}]^j
\int_{{\mathcal X}^{j\beta}_{00}} P^{j\beta}_{00,+} (\de X^0)e^{N(X^0)}.
\end{align}
The absolute convergence of $A_0(z)$ follows from
\begin{lem}
\begin{align}
&(a)\,\,\,\,  \int P^{j\beta}_{00,+}(\de Y^0)e^{N(Y^0)} \leq\frac{1}{2}
e^{j\beta M (e-1)},\nonumber\\
&(b)\,\,\,\, \int P^{j\beta}_{00,+}(\de Y^0)e^{N(Y^0)}N(Y^0) \leq\frac{1}{2}
j\beta e M e^{j\beta M (e-1)},
\nonumber\\
&(c)\,\,\,\, \int P^{j\beta}_{00,+}(\de Y^0)
e^{N(Y^0)}N^2(Y^0) \leq\frac{1}{2}
j\beta e M (   j\beta e M + 1 ) e^{j\beta M (e-1)} \nonumber
\end{align}
\label{Lem1-old-prop1}
\end{lem}
which is proved in the Appendix A1. More precisely
$|A_0(z)| \leq \sum_{j\geq1} q^j(z) <\infty$
if
\begin{equation}
q(z)\equiv z e^{\beta(Me+\Re M_0 + ||\psi||) + 1}<1 .
\label{qz}
\end{equation}
This shows that the pressure $p(z) = p(z,\pi,\psi) = \beta^{-1}A_0(z)$
is analytic on $z$ in (\ref{qz}).

\bigskip

\subsection{The boundary term}

Let us then consider the boundary terms $I_l(R,z),\,\,l=1,\cdots,6$ given by \ref{I1} . We take a face
$\lambda=\lambda_{l_1}\in\Lambda^{(f)}$
 which is a rectangle with sides of length $a_2,\,a_3$ and is
defined by the normal $n_{l_1}$. Next we choose a pair of faces that are adjacent to $\lambda_{l_1}$.
Let $n_{l_2}, n_{l_3}$ be the normals to the chosen faces.
Consider a Cartesian coordinate system with the origin at the point $\lambda_{l_1}\cap\lambda_{l_2}\cap\lambda_{l_3}$
and axes $x_1,x_2,x_3$ along the normals  $ n_{l_1}, n_{l_2}, n_{l_3}$ respectively.

Note that
$\inf\langle(X^0+r,\omega),n_{l_1}\rangle = \inf\langle(X^0+(r_1,0,0),\omega),n_{l_1}\rangle$. This implies
$H(-\inf\langle(X^0+r,\omega),n_{l_1}\rangle -1) = H(-\inf\langle(X^0+(r_1,0,0),\omega),n_{l_1}\rangle -1)$.
Then by translation invariance of the measure $W_{\mu_z}$ and the Ursell function

\begin{align}
I_{l_1}(R,z)= &-
\sum_{r_2=0}^{R a_2}\,\sum_{r_3=0}^{R a_3}\,\sum_{r_1=0}^\infty \sum_{j\geq1} \frac{z^j}{j}
\int_{{\mathcal X}^{j\beta}_{00}} P^{j\beta}_{00} (\de X^0)
e^{-v(X^0)} \int_{\M} W_{\mu_z}(\de \omega) \nonumber\\
& \cdot\frac{\varphi(X^0 +(r_1,0,0),\omega)}{|\omega| +1} H(-\inf\langle(X^0+(r_1,0,0),\omega),n_{l_1}\rangle -1) \nonumber\\
+ &\sum_{r_2=0}^{R a_2}\,\sum_{r_3=0}^{R a_3}\,
\sum_{r_1=Ra_1+1}^\infty \sum_{j\geq1} \frac{z^j}{j}\int_{{\mathcal X}^{j\beta}_{00}} P^{j\beta}_{00}
(\de X^0)e^{-v(X^0)}
 \int_{\M} W_{\mu_z}(\de \omega) \nonumber\\
& \cdot\frac{\varphi(X^0 +(r_1,0,0),\omega)}{|\omega| +1} H(-\inf\langle(X^0+(r_1,0,0),\omega),n_{l_1}\rangle -1) \nonumber\\
=& I'_{l_1}(R,z) + I''_{l_1}(R,z).
\end{align}
Evidently
\begin{equation}
I'_{l_1}(R,z)= R^2 a_2 a_3 A_1(\lambda_{l_1},z)
\end{equation}
where
\begin{align}
A_1(\lambda_{l_1},z)= &A_{l_1}(z)=
-\sum_{r_1=0}^\infty \sum_{j\geq1} \frac{z^j}{j}\int_{{\mathcal X}^{j\beta}_{00}} P^{j\beta}_{00} (\de X^0)
e^{-v(X^0)} \int_{\M} W_{\mu_z}(\de \omega) \nonumber\\
& \cdot\frac{\varphi(X^0 +(r_1,0,0),\omega)}{|\omega| +1} H(-\inf\langle(X^0+(r_1,0,0),\omega),n_{l_1}\rangle -1).
\label{A1}
\end{align}
To prove the absolute convergence of $A_{l_1}(z)$
we note that with the help of the equality
\begin{align}
1_{\M^c(\Lambda)}(X,\omega)= 1(X)1_{\M^c(\Lambda)}(\omega) + 1_{\M(\Lambda)}(\omega) 1_{{\mathcal X}^c(\Lambda)}(X),
\label{decomp-cf}
\end{align}
which holds true for any  $\Lambda\subset \Z^d$,
we have
\begin{align}
H(-\inf\langle(X^0+(r_1,0,0),\omega),n_{l_1}\rangle -1)=1_{\M^c(N_{l_1})}(X,\omega)\nonumber\\
= 1(X)1_{\M^c(N_{l_1})}(\omega) + 1_{\M(N_{l_1})}(\omega) 1_{{\mathcal X}^c(N_{l_1})}(X).
\end{align}
Then   $A_{l_1}(z)$ can be decomposed as:
\begin{align}
A_{l_1}(z)= \widehat{A_{l_1}}(z) + \widetilde{A_{l_1}}(z)
\label{A1hatA1tilda}
\end{align}
where
\begin{align}
\widehat{A_{l_1}}(z)= &
-\sum_{r_1=0}^\infty \sum_{j\geq1} \frac{z^j}{j}\int_{{\mathcal X}^{j\beta}_{00}} P^{j\beta}_{00} (\de X^0)
e^{-v(X^0)} \int_{\M} W_{\mu_z}(\de \omega) \nonumber\\
& \cdot\frac{\varphi(X^0 +(r_1,0,0),\omega)}{|\omega| +1}1_{\M^c(N_{l_1})}(\omega)
\end{align}
and
\begin{align}
\widetilde{A_{l_1}}(z)= &
-\sum_{r_1=0}^\infty \sum_{j\geq1} \frac{z^j}{j}\int_{{\mathcal X}^{j\beta}_{00}} P^{j\beta}_{00} (\de X^0)
e^{-v(X^0)}1_{{\mathcal X}^c(N_{l_1})}(X^0 +(r_1,0,0))
\nonumber\\
& \cdot \int_{\M} W_{\mu_z}(\de \omega) \frac{\varphi(X^0 +(r_1,0,0),\omega)}{|\omega| +1}  1_{\M(N_{l_1})}(\omega).
\end{align}
Let us show first the absolute convergence of $\widehat{A_{l_1}}(z)$.
We shall often use the following inequality
\begin{align}
 \int_{\M} W_{|\mu_z|}(\de \omega) h(\omega)1_{\M^c(\Lambda)}(\omega)\leq \int_{{\mathcal X}} 1_{{\mathcal X}^c(\Lambda)}(X)|\mu_z|
     (\de X )   \int_{\M} W_{|\mu_z|}(\de \omega)h(X,\omega)
 \label{MinlosM-C}
\end{align}
which holds true for all $\Lambda\subset \Z^d$ and any non-negative function
$h(\omega),\,\,\,\omega\in\M$,
The proof follows from  formula (\ref{MM}) and the evident inequality:
$$
1_{\M^c(\Lambda)}(\omega)\leq   \sum _{\bar{\omega}\subset\omega}
1_{\M_1}(\bar{\omega}) 1_{\M^c(\Lambda)}(\bar{\omega})
$$
where $\M_1 = \mathcal X$.

In view of (\ref{MinlosM-C}) and \ref{nsigmaxy} we have
\begin{multline}
\mid\widehat{A_{l_1}}(z)\mid\leq\sum_{r_1=0}^\infty \sum_{j\geq1} \frac{z^j}{j}
\int_{{\mathcal X}^{j\beta}_{00}} \mid P^{j\beta}_{00}\mid (\de X^0) e^{-\Re v(X^0)}
 \int_{\mathcal X} |\mu_z| (\de Y)1_{{\mathcal X}^c(B_r(r_1))}(Y)\\
 \cdot \int_{\M} W_{|\mu_z|}(\de \omega)
 \frac{\mid\varphi(X^0 +(r_1,0,0), Y, \omega)\mid}{|\omega| +2}
\leq\sum_{r_1=0}^\infty \sum_{j\geq1} \frac{z^j}{j}\int_{{\mathcal X}^{j\beta}_{00}}
 \mid P^{j\beta}_{00}\mid (\de X^0)\\
\cdot e^{-\Re v(X^0)} \int_{\mathcal X} |\mu_z| (\de Y)1_{{\mathcal X}^c(B_r(r_1))}(Y)
 |\sigma| (X^0 +(r_1,0,0) ,Y).
\label{A1tp}
\end{multline}

Let
\begin{equation}
q_l(z) \equiv z e^{\beta(M_l e+\Re M_0 + ||\psi||)-1} <1.
\label{qdelta}
\end{equation}
The following result  is the main technical tool for proving the asymptotic expansion of the log-partition function.
 It is proved in Appendix B.
\begin{prop}
\label{MainProp1}
If Assumptions 1,  3 and  Assumption 4 
are fulfilled then for  all $R >0$ and
all $z$ satisfying \ref{qdelta} the following bound holds true
\begin{align}
\sum_{j=1}^\infty \frac{z^j}{j} \int \mid P^{j\beta}_{00}\mid (\de X^0) |e^{-v(X^0)}|
 \int_{{\mathcal X}^c(B_0(R))}\de|\mu_z|(Y) \, |\sigma|(X^0,Y) \nonumber\\
  \leq C(l,p)(1+ \beta e M_l)
 \frac{q_l(z)}{1-q_l(z)}  (1+R)^{-l}.
 \label{boundMainProp1}
\end{align}
where $C(l, p)$ is a positive constant depending on the parameters $l$ and $p$  and $M_l $ is given  by (\ref{Mdelta}).
\end{prop}
We denote all the constants by $C$ indicating only the dependence on the parameters.

With the help of this lemma  it follows from (\ref{A1tp}) that
\begin{equation}
\mid\widehat{A_{l_1}}(z)\mid
\leq  C(l,p) (1+e\beta M_l) \frac{q_l(z)}{1- q_l(z)}  \sum_{r_1=0}^\infty  (1+r_1)^{-l} <\infty
\label{A1hat}
\end{equation}
provided \ref{qdelta}.

Passing to $\widetilde{A_{l_1}}(z)$ and using formulas \ref{intdomega}, (\ref{Rev-a-b-norm}) we find
\begin{multline}
\mid\widetilde{A_{l_1}}(z)\mid
\leq
\sum_{r_1=0}^\infty \sum_{j\geq1} \frac{(z e^{\beta(M+\Re M_0 + ||\psi||)+1})^j}{j}
\int_{{\mathcal X}^{j\beta}_{00}} P^{j\beta}_{00,+} (\de X^0)
 e^{N(X^0)}\\
\cdot 1_{\sup_t |X^0(t)|>r_1 }(X^0).
 \label{A1tilda}
 \end{multline}
To estimate integrals over large loops, like the one in \ref{A1tilda}, we use the following
\begin{lem}
\label{Lemma2}
If the potential $\pi$ satisfies the condition (\ref{Mdelta})
then for any $R>0$ and all $j=1,2,\cdots,$
\begin{align}
&(a)\,\,
\int P^{j\beta}_{00,+}(\de X^0)e^{N(X^0)}
1_{\sup_t |X^0(t)|\geq R}(X^0)
\leq
e^{j\beta (M_l e-M)} (1+R)^{-l},\nonumber\\
&(b) \,\, \int P^{j\beta}_{00,+}(\de X^0)
e^{N(X^0)}N(X^0)1_{\sup_t |X^0(t)|\geq R}(X^0)
 \leq j\beta e M_l e^{j\beta (M_l e-M)}
  (1+R)^{-l},  \nonumber\\
&(c)\,\, \int P^{j\beta}_{00,+}(\de X^0)
e^{N(X^0)}N^2(X^0)1_{\sup_t |X^0(t)|\geq R}(X^0)
\leq \nonumber
 j\beta e M_l ( j\beta e M_l +1)
 \nonumber \\
 &\hspace{9cm}
 \times  e^{j\beta (M_l e-M)}(1+R)^{-l}
  \nonumber
\end{align}
with $M_l$ defined by  \ref{Mdelta}.
\end{lem}

By Lemma \ref{Lemma2} (a) it follows from \ref{A1tilda} that
\begin{equation}
\mid\widetilde{A_{l_1}}(z)\mid
< \frac{q_l(z)}{1-q_l(z)} \sum_{r_1=0}^\infty  (1+r_1)^{-l}<\infty
 \label{estA1tilde}
 \end{equation}
 for all $z$ satisfying (\ref{qdelta}).
Thus combining (\ref{A1hatA1tilda}), (\ref{A1hat}) and (\ref{A1tilda}) we get
the absolute convergence of $A_{l_1}(z)$ and the analyticity on $z$ in (\ref{qdelta}).

Under the same condition (\ref{qdelta}) the term  $I''_{l_1}(R,z)$ can be estimated as
\begin{equation}
I''_{l_1}(R,z)
\leq
[ C(l,p) (1+e\beta M_l)+1]
 \frac{q_\delta(z)}{1- q_l(z)}
 \, a_2a_3 R^2  \\
\sum_{r_1=R a_1}^\infty  (1+r_1)^{-l}
\end{equation}
that goes to $0$ as $R\to \infty$ since $l>3$.
This implies
\begin{equation}
I_{l_1}(R,z) = R^2 \, a_2a_3 \, A_{l_1}(z) + o(1)  =  R^2 |\lambda_{l_1}| A_1(\lambda_{l_1},z) + o(1),\,\,\,  \textrm{as}\,\, R\rightarrow\infty
\label{I1Final}
\end{equation}
provided (\ref{qdelta}).

Thus the boundary term of the geometric expansion of $\ln Z(\Lambda_R, z)$ is
\begin{equation}
 R^2 \sum_{\lambda \in \Lambda^{(f)}}|\lambda| \, A_1(\lambda,z).
\label{boundterm}
\end{equation}

\subsection{The edge terms}

The contribution to this (one-dimensional) term is coming from the quantities
$I_{l_1,l_2}(R,z)$ such that $\lambda_{l_1}$ and $\lambda_{l_2}$ are adjacent faces.
We consider a face $\lambda_{l_3}$ which is adjacent to both faces $\lambda_{l_1}$ and $\lambda_{l_2}$.
Assume that the edge $\lambda_{l_1,l_2}$ has the length $a_3$ and
and define a Cartesian coordinate system with the origin at the point $\lambda_{l_1}\cap\lambda_{l_2}\cap\lambda_{l_3}$
and axes $x_1,x_2,x_3$ along the normals  $ n_{l_1}, n_{l_2}, n_{l_3}$ respectively.

Using the evident equality
\begin{equation}
\prod_{i=1,2} H(-\inf\langle(X^0+r,\omega),n_{l_i}\rangle -1)
 =
\prod_{i=1,2} H(-\inf\langle(X^0+(r_1,r_2,0),\omega),n_{l_i}\rangle -1),
\end{equation}
the translation invariance arguments  we
find from (\ref{I12}) that
\begin{multline}
I_{l_1,l_2}(R,z)
=\sum_{r_1=0}^{\infty}\,\sum_{r_2=0}^{\infty}
\,
\sum_{r_3=0}^{R a_3}
 \sum_{j\geq1}\frac{z^j}{j}
  \int_{{\mathcal X}^{j\beta}_{00}} P^{j\beta}_{00} (\de X^0)e^{-v(X^0)}
 \int_{\M} W_{\mu_z}(\de \omega)\\
\cdot \frac{\varphi(X^0 +(r_1,r_2,0),\omega)}{|\omega| +1}
\prod_{i=1,2} H(-\inf\langle(X^0+(r_1,r_2,0),\omega),n_{l_i}\rangle -1)
\\
-  \sum_{ (r_1,r_2)\in \Lambda^c_{3}(R)}\,
\sum_{r_3=0}^{R a_3} \sum_{j\geq1}\frac{z^j}{j}
  \int_{{\mathcal X}^{j\beta}_{00}} P^{j\beta}_{00} (\de X^0)e^{-v(X^0)}
 \int_{\M} W_{\mu_z}(\de \omega)\\
\cdot \frac{\varphi(X^0 +(r_1,r_2,0),\omega)}{|\omega| +1}
\prod_{i=1,2} H(-\inf\langle(X^0+(r_1,r_2,0),\omega),n_{l_i}\rangle -1)
 \\
  \equiv I'_{l_1,l_2}(z) - I''_{l_1,l_2}(R,z)
\label{I'I''12}
\end{multline}
where
$\Lambda^c_{3}(R) =
  \{
  (r_1,r_2,0)\,\mid \, r_1,r_2\geq 0)
   \}
   \setminus
   \Lambda^{(f)}_{3}(R)
   $
   is the complement of the face
   $\Lambda^{(f)}_{3}(R) =
 \{(r_1,r_2,0)\,\mid\,0\leq r_1\leq Ra_1 ;  0\leq r_2\leq Ra_2 \}
   $ with respect to the corresponding quadrant of the plane $r_3=0$.
   Then,
   \begin{equation}
   I'_{l_1,l_2}(z)=R a_3  A_2(\lambda_{l_1,l_2}, z)
   \end{equation}
   where
   \begin{multline}
 A_2(\lambda_{l_1,l_2}, z) =  A_{l_1,l_2}(z)
    =\sum_{r_1=0}^{\infty}\,\sum_{r_2=0}^{\infty}
    \sum_{j\geq1}\frac{z^j}{j}
  \int_{{\mathcal X}^{j\beta}_{00}} P^{j\beta}_{00} (\de X^0)e^{-v(X^0)}
 \int_{\M} W_{\mu_z}(\de \omega)
 \\
\cdot \frac{\varphi(X^0 +(r_1,r_2,0),\omega)}{|\omega| +1}
\cdot \,
\prod_{i=1,2} H(-\inf\langle(X^0+(r_1,r_2,0),\omega),n_{l_i}\rangle -1)
\label{A12}
   \end{multline}

Let us show the absolute convergence of $ A_{l_1,l_2}(z)$.
With the help of the equation (\ref{decomp-cf}) we have
\begin{multline}
\prod_{i=1,2} H(-\inf\langle(X^0+(r_1,r_2,0),\omega),n_{l_i}\rangle -1)
\leq 1_{\M^c(B_{(r_1,r_2,0)}(\max(r_1,r_2)))}(\omega)\\
 +  1_{{\mathcal X}^c(B_{(r_1,r_2,0)}(\max(r_1,r_2)))}(X).
\end{multline}
Therefore
\begin{multline}
|A_{l_1,l_2}(z)|\leq
  \sum_{r_1=0}^{\infty}\,\sum_{r_2=0}^{\infty}\, \sum_{j\geq1}\frac{z^j}{j}
  \int_{{\mathcal X}^{j\beta}_{00}} P^{j\beta}_{00} (\de X^0)e^{-v(X^0)}
 \int_{\M} W_{\mu_z}(\de \omega)\\
\cdot \frac{\varphi(X^0 +(r_1,r_2,0),\omega)}{|\omega| +1}
 1_{\M^c(B_{(r_1,r_2,0)}(\max(r_1,r_2)))}(\omega)
 + \sum_{r_1=0}^{\infty}\,\sum_{r_2=0}^{\infty}\, \sum_{j\geq1}\frac{z^j}{j}\\
 \cdot \int_{{\mathcal X}^{j\beta}_{00}} P^{j\beta}_{00} (\de X^0)e^{-v(X^0)}
 1_{{\mathcal X}^c(B_{(r_1,r_2,0)}(\max(r_1,r_2)))}   (X^0 + (r_1,r_2,0))\\
\cdot \int_{\M} W_{\mu_z}(\de \omega)\frac{\varphi(X^0 +(r_1,r_2,0),\omega)}{|\omega| +1}
\equiv \widehat{A_{l_1,l_2}}(z) + \widetilde{A_{l_1,l_2}}(z).
\label{A12Decomp}
\end{multline}
Using first (\ref{MinlosM-C}) then Proposition \ref{MainProp1}
we can estimate $ \widehat{A_{l_1,l_2}}(z)$ as
\begin{multline}
\mid\widehat{A_{l_1,l_2}}(z)\mid
\leq
  \sum_{r_1=0}^{\infty}\,\sum_{r_2=0}^{\infty}\, \sum_{j\geq1}\frac{z^j}{j}
  \int_{{\mathcal X}^{j\beta}_{00}} |P^{j\beta}_{00}| (\de X^0)e^{-\Re v(X^0)}
 \int_{\mathcal X} |\mu_z|(\de Y) \\
 \cdot1_{{\mathcal X}^c(B_{(r_1,r_2,0)}(\max(r_1,r_2))} (Y)
 |\sigma| (X^0 +(r_1,r_2,0),Y)
 \\
 \leq C(l,p) (1+\beta e M_{l})
  \frac{q_{l}}{1-q_{l}}
 \cdot\sum_{r_1=0}^{\infty}\,\sum_{r_2=0}^{\infty}\,(1+\max(r_1,r_2))^{-l}
 \label{A12hat}
 \end{multline}
with an absolute convergent double sum in the last line of \ref{A12hat}:
\begin{equation}
\sum_{r_1=0}^{\infty}\,\sum_{r_2=0}^{\infty}\,(1+\max(r_1,r_2))^{-l}
=\sum_{s=0}^\infty (2s+1) (1+s)^{-l}
<\infty.
 \end{equation}

 Passing to $\widetilde{A_{l_1,l_2}}(z)$ and using formulas \ref{intdomega},
 (\ref{Rev-a-b-norm}) and Lemma \ref{Lemma2} (a) we have that for all $z$ satisfying (\ref{qdelta})
\begin{multline}
\mid\widetilde{A_{l_1,l_2}}(z)\mid\leq
\sum_{r_1=0}^\infty
\sum_{r_2=0}^{\infty} \sum_{j\geq1} \frac{(z e^{\beta(M+\Re M_0 + ||\psi||)+1})^j}{j}
\int_{{\mathcal X}^{j\beta}_{00}} P^{j\beta}_{00,+} (\de X^0)
 e^{N(X^0)} \\
\cdot1_{\sup_t |X^0(t)|>\max(r_1,r_2)}(X^0)
 \leq \frac{q_l(z)}{1-q_l(z)} \sum_{r_1=0}^\infty
  \sum_{r_1=0}^\infty (1+\max \{ r_1,r_2\})^{-l}.
  \label{A12tilda}
 \end{multline}
Thus combining (\ref{A12Decomp}), (\ref{A12hat}) and (\ref{A12tilda}) we get   the absolute convergence of $A_{l_1,l_2}(z)$
under the condition (\ref{qdelta}).

 The term  $I''_{l_1,l_2}(R,z)$ can be estimated as
\begin{equation}
I''_{l_1,l_2}(R,z)
\leq C\, R \sum_{s\geq R \min (a_1,a_2)}
(2s+1)(1+s)^{-l} \underset{R\to \infty}{ \longrightarrow }0.
\end{equation}

 This implies
   \begin{equation}
   I_{l_1,l_2}(R, z)=R a_3  A_{l_1,l_2}(z) + o(1) = R |\lambda_{l_1,l_2}| A_2(\lambda_{l_1,l_2},z) + o(1), \,\,\,\textrm{as}\,\, R\rightarrow\infty
  \label{I12Final}
   \end{equation}
with     $A_{l_1,l_2}$ given by \ref{A12}.
Thus the edge term of the geometric expansion is
 \begin{equation}
   R \sum_{\lambda \in \Lambda^{(e)}} |\lambda|  A_2(\lambda,z).
  \label{edge term}
 \end{equation}

\subsection{The corner terms}

We consider now the next  (constant) term  of the expansion.
The contribution to this term which does not depend on $R$ is coming from the quantities
$I_{l_1,l_2,l_3}(R,z)$ given by (\ref{I123}) for which the corresponding faces are adjacent.
We take a vertex (corner) $\lambda = \lambda_{l_1,l_2,l_3}\in \Lambda^{(v)}$
which is defined by three adjacent faces $\lambda_{l_1}, \lambda_{l_2}, \lambda_{l_3}$
with the normals $n_{l_1}, n_{l_2}, n_{l_3}$. As before we
 define a Cartesian coordinate system with the origin at the point $\lambda = \lambda_{l_1}\cap \lambda_{l_2}\cap \lambda_{l_3}$
 and axes along the normals  $ n_{l_1}, n_{l_2}, n_{l_3}$.
Then it follows from (\ref{I123}) that
\begin{multline}
I_{l_1,l_2,l_3}(R,z)=
-  \sum_{r_1=0}^{\infty}\,\sum_{r_2=0}^{\infty}\,
\sum_{r_3=0}^{\infty}
\sum_{j\geq1}\frac{z^j}{j}
  \int_{{\mathcal X}^{j\beta}_{00}} P^{j\beta}_{00} (\de X^0)e^{-v(X^0)}
 \int_{\M} W_{\mu_z}(\de \omega)\\
\cdot \frac{\varphi(X^0 +r,\omega)}{|\omega| +1}
\prod_{i=1,2,3} H(-\inf\langle(X^0+r,\omega),n_{l_i}\rangle -1)\\
+  \sum_{ r\in \mathbb{Z}^3_+  \setminus \Lambda (R)}\,
 \sum_{j\geq1}\frac{z^j}{j}
 \int_{{\mathcal X}^{j\beta}_{00}} P^{j\beta}_{00} (\de X^0)e^{-v(X^0)}
 \int_{\M} W_{\mu_z}(\de \omega)\\
\prod_{i=1,2,3} H(-\inf\langle(X^0+r,\omega),n_{l_i}\rangle -1)\\
  \equiv I'_{l_1,l_2,l_3}(z) + I''_{l_1,l_2,l_3}(R,z)
\end{multline}
where  $\mathbb{Z}^3_+ =\{ (r_1,r_2,r_3)\in \mathbb{Z}^3 \,\mid \, r_1,r_2, r_3\geq 0   \} $.

We need to show the absolute convergence of $I_{l_1,l_2,l_3}'(z)$.
Setting  $r=(r_1,r_2,r_3)$ with the help of the equality (\ref{decomp-cf})
 we have:
\begin{multline}
\prod_{i=1,2,3} H(-\inf\langle(X^0+r,\omega),n_{l_i}\rangle -1)
\leq
1_{\M^c (B_r(\max(r_1,r_2,r_3)))}(\omega)\nonumber\\
 + 1_{{\mathcal X}^c (B_r(\max(r_1,r_2,r_3)))}(X^0 + r).
\end{multline}
Therefore
\begin{multline}
|I_{l_1,l_2,l_3}'(z)|\leq
  \sum_{r_1=0}^{\infty}\,\sum_{r_2=0}^{\infty}\,
  \sum_{r_3=0}^{\infty}
   \sum_{j\geq1}\frac{z^j}{j}
  \int_{{\mathcal X}^{j\beta}_{00}} |P^{j\beta}_{00}| (\de X^0) |e^{-v(X^0)}|
 \int_{\M} W_{|\mu_z|}(\de \omega)\\
\cdot \frac{|\varphi(X^0 +r,\omega)|}{|\omega| +1}
\Bigl[ 1_{\M^c (B_r(\max(r_1,r_2,r_3)))}(\omega)
+ 1_{{\mathcal X}^c (B_r(\max(r_1,r_2,r_3)))}(X^0 + r)\Bigr].
\label{I'123}
\end{multline}
The term of (\ref{I'123}) which correspond to the first summand in the brackets can be estimated similarly to
$\widehat{A_{l_1,l_2}}(z)$. Treating the term corresponding to the second summand in the brackets by means of
 (\ref{intdomega}), (\ref{Rev-a-b-norm}) and
 Lemma \ref{Lemma2} 
 we get the absolute convergence
 of $I_{l_1,l_2,l_3}'(z)$ for all $z$ satisfying (\ref{qdelta}):
\begin{multline}
|I_{l_1,l_2,l_3}'(z)|\leq
\bigl[ C(l,p) (1+\beta e M_{l}) + 1 \bigr]
\frac{q_l(z)}{1-q_l(z)} \sum_{r_1=0}^\infty
  \sum_{r_2=0}^\infty \\
\cdot   \sum_{r_3=0}^\infty (1+\max(r_1,r_2,r_3))^{-l} <\infty
  \end{multline}
Here we used the fact that
\begin{equation}
\sum_{r_1=0}^{\infty}\,\sum_{r_2=0}^{\infty}\,
\sum_{r_3=0}^{\infty}\,(1+\max(r_1,r_2,r_3))^{-l}
\leq C\sum_{s=0}^\infty  (1+s)^{-l+2}
<\infty.
 \end{equation}

The term  $I''_{l_1,l_2,l_3}(R,z)$ can be estimated as
\begin{equation}
I''_{l_1,l_2,l_3}(R,z)
\leq C \sum_{s\geq R \min (a_1,a_2,a_3)}
(1+s)^{-l+2}
 \underset{R\to \infty}{ \longrightarrow }0.
\end{equation}
Hence
\begin{equation}
I_{l_1,l_2,l_3}(R,z)= I'_{l_1,l_2,l_3}(z) +o(1) \,\,\textrm{as}\,\, R\rightarrow\infty
\end{equation}
where
\begin{multline}
 I'_{l_1,l_2,l_3}(z) \equiv A_3(\lambda_{l_1,l_2,l_3}, z) =
-  \sum_{r_1=0}^{\infty}\,\sum_{r_2=0}^{\infty}\,
\sum_{r_3=0}^{\infty}
\sum_{j\geq1}\frac{z^j}{j}
  \int_{{\mathcal X}^{j\beta}_{00}} P^{j\beta}_{00} (\de X^0)
  e^{-v(X^0)}\\
\cdot \int_{\M} W_{\mu_z}(\de \omega)
 \frac{\varphi(X^0 +r,\omega)}{|\omega| +1}
\prod_{i=1,2,3} H(-\inf\langle(X^0+r,\omega),n_{l_i}\rangle -1).
\label{I123Final}
\end{multline}

Then the constant term of the expansion can be written as
\begin{equation}
\sum_{\lambda \in \Lambda^{(v)}} A_3(\lambda,z)
\label{constterm}
\end{equation}
with $A_3(\lambda,z) = I'_{l_1,l_2,l_3}(z) $ if $\lambda = \lambda_{l_1}\cap\lambda_{l_2}\cap\lambda_{l_3} $.

\subsection{The reminder}

It remains to consider  the reminder term $Q(R,z)$ of the expansion.
The contribution to this term is coming from the quantities $I_{L^*}$ with those
$L^*\subset \{1,2,\dots,6\}$ which contain at least two indices, say $l_1, l_2$, such that the corresponding faces
are parallel.
For such $L^*$, with the help of the equality (\ref{decomp-cf}),
 we have:
\begin{equation}
\prod_{l\in L^*} H(-\inf\langle(X^0+r,\omega),n_{i}\rangle -1)
 \leq 1_{\M^c(B_r(\frac{R a}{3}))}(\omega)  +  1_{{\mathcal X}^c(B_r(\frac{R a}{3}))}(X^0+r)
 \nonumber
\end{equation}
where $a=\min(a_1,a_2,a_3)$.
Therefore it follows from (\ref{Itrash}) that
\begin{multline}
|I_{L^*}(R,z)|\leq
 \sum_{r\in\Lambda_R}
   \sum_{j\geq1}\frac{z^j}{j}
  \int_{{\mathcal X}^{j\beta}_{00}} |P^{j\beta}_{00}| (\de X^0)|e^{-v(X^0)}|
 \int_{\M} W_{|\mu_z|}(\de \omega)\\
\cdot \frac{|\varphi(X^0 +r,\omega)|}{|\omega| +1}
 \Bigl[ 1_{\M^c(B_r(\frac{R a}{3}))}(\omega)
+  1_{{\mathcal X}^c(B_r(\frac{R a}{3}))}(X^0+r)  \Bigr]
\end{multline}
Using again (\ref{MinlosM-C}) and Proposition \ref{MainProp1} we have
\begin{multline}
|I_{L^*}(R,z)|\leq
C(l,p) (2+\beta e M_{l})
\frac{q_{l}}{1-q_{l}}
R^3 a_1 a_2 a_3 (1+R)^{-l}
\underset{R\to \infty}{ \longrightarrow }0
 \label{Ihat1234}
 \end{multline}
since  $l>3$.
This in view of (\ref{QRz})  implies
\begin{equation}
Q(R,z) = o(1),\,\,\,R\rightarrow \infty.
\label{trash}
\end{equation}

Collecting all the non decreasing terms of the expansion given by (\ref{A0}), (\ref{A1}), (\ref{A12}), and (\ref{I123Final}) and taking into account
(\ref{trash}) we arrive to the final expansion
\begin{align}
\ln Z(\Lambda_R ,z)&= R^3 |\Lambda | A_0 (z)
+  R^2 \sum_{\lambda\in\Lambda^{(f)}} |\lambda | A_1 (\lambda, z)
+  R \sideset{}{^{(\text{adj})}}\sum_{\lambda\in\Lambda^{(e)} } |\lambda | A_2(\lambda,z) \nonumber
\\
&+  \sideset{}{^{(\text{adj})}} \sum_{\lambda \in \Lambda^{(v)}}  A_3(\lambda, z)
+o(1).
\end{align}
 If in addition  the transverse and longitudinal potentials  are invariant with respect to the automorphism group of the lattice $\Z^3$ then
\begin{align}
\ln Z(\Lambda_R ,z)&= R^3 |\Lambda | A_0 (z)
+  R^2  |\partial\Lambda | A_1 ( z)
+  R \sideset{}{^{(\text{adj})}}\sum_{\lambda\in\Lambda^{(e)} } |\lambda | A_2(z) \nonumber
\\
&+  8  A_3( z)
+o(1).
\end{align}
Here
\begin{align}
A_1(z)= &
-\sum_{r_1=0}^\infty \sum_{j\geq1} \frac{z^j}{j}\int_{{\mathcal X}^{j\beta}_{00}} P^{j\beta}_{00} (\de X^0)
e^{-v(X^0)} \int_{\M} W_{\mu_z}(\de \omega) \nonumber\\
& \cdot\frac{\varphi(X^0 +(r_1,0,0),\omega)}{|\omega| +1} H(-\inf\langle(X^0+(r_1,0,0),\omega),e_1\rangle -1),
\label{A^b}
\end{align}
\begin{multline}
A_2(z)
 =\sum_{r_1=0}^{\infty}\,\sum_{r_2=0}^{\infty}
  \sum_{j\geq1}\frac{z^j}{j}
  \int_{{\mathcal X}^{j\beta}_{00}} P^{j\beta}_{00} (\de X^0)e^{-v(X^0)}
 \int_{\M} W_{\mu_z}(\de \omega)
 \\
\cdot \frac{\varphi(X^0 +(r_1,r_2,0),\omega)}{|\omega| +1}
\cdot \,
\prod_{i=1,2} H(-\inf\langle(X^0+(r_1,r_2,0),\omega),e_{i}\rangle -1)
\label{A^e}
\end{multline}
and
\begin{multline}
A_3(z) =
- \sum_{r_1=0}^{\infty}\,\sum_{r_2=0}^{\infty}\,
\sum_{r_3=0}^{\infty}
\sum_{j\geq1}\frac{z^j}{j}
  \int_{{\mathcal X}^{j\beta}_{00}} P^{j\beta}_{00} (\de X^0)e^{-v(X^0)}
 \int_{\M} W_{\mu_z}(\de \omega)\\
\cdot \frac{\varphi(X^0 +r,\omega)}{|\omega| +1}
\prod_{i=1,2,3} H(-\inf\langle(X^0+r,\omega),e_{i}\rangle -1).
\label{A^v}
\end{multline}

\emph{Acknowledgements.} The second author wishes to thank the Centre de Physique Th\'eorique  Marseille for
 kind hospitality and to acknowledge the Universit\'e du Sud Toulon Var for financial support.

\subsection*{Appendix A1: proofs of Lemma \ref{Lem1-old-prop1} and Lemma \ref{Lemma2} }

\textbf{Proof of Lemma \ref{Lem1-old-prop1}.}

We note that $\sup_{r\in\Z^d, r\neq 0}\frac{\pi(r)}{2M}\leq \frac{1}{2}$
hence
\begin{equation}
 \sumtwo{r_1,\cdots,r_n}{r_1+\cdots+r_n=0} \prod_{i=1}^n\frac{|\pi(r_i)|}{2M}
 \leq\frac{1}{2}.
 \label{boundsumpi}
\end{equation}
Therefore
\begin{eqnarray}
\int P^{j\beta}_{00,+}(\de Y^0)e^{N(Y^0)} =e^{-j\beta M} \sum_{n=0}^\infty \frac{(j\beta M e)^n}{n!}
 \sumtwo{r_1,\cdots,r_n}{r_1+\cdots+r_n=0} \prod_{i=1}^n\frac{|\pi(r_i)|}{2M}\nonumber.
\end{eqnarray}
This implies  (a), equations (b) and (c) can be proved in the same way. Lemma \ref{Lem1-old-prop1} is proved.

\textbf{Proof of Lemma \ref{Lemma2}.}
Let us prove the equation (c).
We note that for any loop $X^0$ which satisfies $\sup_t |X^0(t)|\geq R$  and has the jumps $r_1 ,\cdots , r_n$, 
there exists a number $k,\,1\leq k\leq n$ such that
$|r_1 +\cdots + r_k| \geq R$.
Hence in view of (\ref{Mdelta}) we have
\begin{align}
 \int  P^{j\beta}_{00,+} &(\de X^0) 1_{\sup_t |X^0(t)|\geq R}(X^0)e^{N(X^0)}  N^2(X^0) \nonumber \\
 & = e^{-j\beta M}\sum_{n=0}^\infty \frac{(j\beta M)^n}{n!} n^2 e^n
 \cdot \sumtwo{r_1 +\cdots + r_n=0}{|r_1 +\cdots + r_k| \geq R}
 \prod_{i=1}^n \frac{|\pi(r_i)|}{2M} \nonumber \\
& \leq e^{-j\beta M} j\beta M e \sum_{n=1}^\infty \frac{(j\beta M e)^{n-1}}{(n-1)!}n
 \cdot\left[ \sum_{r\in \Z^3, r\neq 0} \frac{|\pi(r)|(1+r)^l}{2M} \right]^n (1+R)^{-l}
  \nonumber \\
 &\leq j\beta e M_l
 ( j\beta e M_l  +1)
 e^{j \beta( M_{l} e -M)} (1+R)^{-l}.
\end{align}
The equations (a) and (b) can be proved in the same way.
This ends the proof of Lemma \ref{Lemma2}.

\subsection*{Appendix A2: proof of Assumption 1}

First let us see that $b(X)=\frac{1}{2}\Re v(X) + \frac{1}{2}\beta ||\psi||\,|X|\geq0$ for all  admissible composite loops $X\in\mathcal{X} $.
Indeed
\begin{multline}
\mid \Re v(X)\mid
\leq \frac{1}{2} \int_0^\beta \sum_{x\in X}\sumtwo{r\in\Z^d}{r\neq x(t)}  \mid\psi(x(t)-r)\mid\de t
\leq \beta ||\psi|| \mid X \mid.
\label{Reof v}
\end{multline}
 Here in the first inequality we used the fact that $X$ is an admissible composite loop.

Next we  check the Assumption 1 itself. We need to prove that for any $n$ and all $X_1,\cdots,X_n \in\mathcal X$,
\begin{equation}
\sum_{1\leq i<j \leq n} \Re u(X_i,X_j) \geq -\frac{1}{2}\left[\sum_{i=1}^n  \Re v(X_i) +\beta ||\psi|| \sum_{i=1}^n \mid X_i \mid \right].
\label{Ass1}
\end{equation}
Let $E(X_1,\cdots,X_n)= \{x\in\mathcal{X}\,\mid\, \exists i,\,\,1\leq i\leq n; \,x\in X_i   \}$
be the family of all elementary constituents of the set of composite loops $X_1,\cdots,X_n$.
We have
\begin{multline}
\mid\sum_{1\leq i<j \leq n} \Re u(X_i,X_j) +\frac{1}{2}\sum_{i=1}^n  \Re v(X_i) \mid
\leq \sum_{1\leq i<j \leq n} \mid u(X_i,X_j)\mid\\
+\sum_{i=1}^n \mid  v(X_i)\mid
\leq \frac{1}{2}\int_0^\beta \de t \sum_{x\in E(X_1,\cdots,X_n)}
 \sum_{r\in\Z^d,r\neq x(t)} |\psi (x(t)-r)|\\
 \leq\frac{1}{2}\beta ||\psi|| \sum_{i=1}^n |X_i|.
\end{multline}
This proves \ref{Ass1}.

\subsection*{Appendix A3: proof of Assumption 3}

The next result gives the condition under which Assumption 3 (hence also Assumption 2) is valid.
\begin{prop} If $z$ satisfies the condition
\begin{equation}
p(z)=   C(\beta)
\sum_{j\geq1} j [z e^{\beta(Me +\Re M_0 + 2||\psi||)+1}]^{j} <1
\label{condpz}
\end{equation}
where
\begin{equation}
 C(\beta)= 2^{-1}[1+3\beta e M + (\beta e M)^2 + \beta ||\psi||(1+ \beta e M)].
\label{Const.inAss3}
\end{equation}
then the Assumption 3 holds true with $a(X)= N(X) + |X|,\,\,b(X)= \frac{1}{2} \Re  v(X) +\frac{1}{2}\beta ||\psi|| | X |$
and $p=p(z)$ given by \ref{condpz}.
\label{Prop1Ass3}
\end{prop}
\textbf{Proof.}
Let
\begin{equation}
J(X,z)\doteqdot\int_{\mathcal{X}}\mid \mu_z \mid (\de Y) \mid \zeta(X,Y)\mid e^{a(Y)+2b(Y)} a(Y).
\end{equation}
Using  translation invariance of the functions $a,b$ and the measure  $ P^{j\beta}_{rr}$
as well as Fubini's theorem and formula (\ref{Rev-a-b-norm}) we can write
\begin{multline}
J(X,z)=
\sum_{j\geq1}\frac{[z e^{\beta(M+\Re M_0 + ||\psi||)+1}]^{j}}{j}
\int P^{j\beta}_{00,+}(\de Y^0) e^{N(Y^0)}a(Y^0)\\
\cdot\sum_{r\in\Z^d}\mid \zeta(X,Y^0+r)\mid.
\end{multline}
Let
\begin{equation}
T(X,Y)=\{r\in\Z^d\,\mid \,\exists t\in(0,\beta),\,x\in X, y\in Y \textrm{such that}\, r=x(t)-y(t)\}.
\end{equation}
Then putting $T^c =\Z^d\setminus T$ we have
\begin{eqnarray}
J(X,z)
=\sum_{j\geq1}\frac{[z e^{\beta(M+\Re M_0 + ||\psi||)+1}]^{j}}{j}
\int P^{j\beta}_{00,+}(\de Y^0) e^{N(Y^0)}a(Y^0) \nonumber\\
\cdot\sum_{r\in T(X,Y^0)}\mid  \zeta(X,Y^0+r)\mid +\sum_{j\geq1}\frac{[z e^{\beta(M+\Re M_0 + ||\psi||)+1}]^{j}}{j}
\int P^{j\beta}_{00,+}(\de Y^0) \nonumber\\
\cdot e^{N(Y^0)}  a(Y^0) \sum_{r\in T^c(X,Y^0)}\mid \zeta(X,Y^0+r)\mid \equiv J'(X,z) + J''(X,z).
\label{Ass3J(X,z)}
\end{eqnarray}

Note that $\mid \zeta(X,Y^0+r)\mid =1$ for all $r\in T(X,Y^0)$. Hence
\begin{eqnarray}
 J'(X,z)
\leq\sum_{j\geq1}\frac{[z e^{\beta(M+\Re M_0 + ||\psi||)+1}]^{j}}{j}
\int P^{j\beta}_{00,+}(\de Y^0) e^{N(Y^0)}a(Y^0)\nonumber\\
\cdot\mid T(X,Y^0)\mid .
\label{InegA}
\end{eqnarray}
To estimate $\mid T(X,Y^0)\mid$ we note that for any two elementary (of length $\beta$) paths $x$ and $y$, $|T(x,y)|\leq  N(x)
+ N(y)+1$.
Hence
\begin{equation}
|T(X,Y^0)|\leq |Y^0| ( N(X) + |X|) + |X| N(Y^0) = a(X)a(Y^0).
\label{InegT}
\end{equation}
Substituting (\ref{InegT}) into (\ref{InegA})
 we get
\begin{equation}
 J'(X,z)
\leq a(X)
\sum_{j\geq1}\frac{\left[z e^{\beta(M+\Re M_0 + ||\psi||)+1}\right]^{j}}{j}
\int P^{j\beta}_{00,+}(\de Y^0) e^{ N(Y^0)} a^2(Y^0).
\end{equation}
Since by formula (\ref{onlyjumps}) and Lemma \ref{Lem1-old-prop1}
\begin{equation}
\int P^{j\beta}_{00,+}(\de Y^0) e^{ N(Y^0)} a^2(Y^0)
\leq \frac{j^2}{2}[1+3\beta e M + (\beta e M )^2] e^{j\beta M (e-1)}
\end{equation}
we find that
\begin{equation}
 J'(X,z)
 \leq a(X) \frac{1}{2}[1+3\beta e M + (\beta e M )^2] \sum_{j\geq1}  j q^j(z)
\label{boundJ'}
\end{equation}
where $q(z)$ is given by (\ref{qz}).

To treat the term $ J''(X,z)$ from \ref{Ass3J(X,z)} we use the following formula
\begin{equation}
\sum_{r\in T^c(X,Y^0)}   \mid \zeta(X,Y^0+r)\mid \leq |X| \beta ||\psi|| |Y^0|  e^{\beta ||\psi|| |Y^0|}
\label{Lem2-old-prop2}
\end{equation}
which holds true for  all admissible $X, Y \in \mathcal X$. Indeed
\begin{equation}
  \mid e^{- u(X,Y^0+r)} -1\mid \leq  |u(X,Y^0+r)| e^{\beta ||\psi|| |Y^0|}.
  \label{estmodMayer}
\end{equation}
Then (\ref{Lem2-old-prop2}) follows from
\begin{multline}
\sum_{r\in T^c(X,Y^0)}   \mid \zeta(X,Y^0+r)\mid\leq e^{\beta ||\psi|| |Y^0|}
 \int_0^\beta \de t \sum_{x\in X}\sum_{y^0\in Y^0}\\
\cdot \sum_{r\in T^c(X,Y^0)} |\psi(x(t)-y^0(t)-r)|.
\end{multline}
On the other hand
 Lemma \ref{Lem1-old-prop1} implies
\begin{equation}
\int P^{j\beta}_{00,+}(\de Y^0)
 e^{N(Y^0)}  a(Y^0)\leq
 \frac{1}{2}j(1+\beta M e) e^{j\beta(Me-1)}.
\label{inta(Y^0)}
\end{equation}
Then applying first (\ref{Lem2-old-prop2}) then (\ref{inta(Y^0)}) we find that
\begin{equation}
 J''(X,z)
\leq a(X) \frac{1}{2}\beta ||\psi||(1+ \beta e M) \sum_{j\geq1}j [q(z)e^{\beta ||\psi||}]^{j}    .
\label{boundJ''}
\end{equation}
Finally combining \ref{boundJ'}, \ref{boundJ''} and \ref{Ass3J(X,z)} we get
\begin{equation}
 J(X,z)  \leq  a(X) C(\beta)
\sum_{j\geq1}j [q(z)e^{\beta ||\psi||}]^{j}
\label{Jfinal}
\end{equation}
where $C(\beta)$ is given by \ref{Const.inAss3}.
Proposition \ref{Prop1Ass3} is proved.

\section*{Appendix A4: proof of Assumption 4}

Here we find the conditions under which Assumption 4 holds true. They are given in Proposition \ref{prop4-oldprop5} below.
\begin{prop}
\label{prop4-oldprop5}
Let the transverse and longitudinal potentials $\pi$ and $\psi$ satisfy respectively the conditions (\ref{Mdelta})
and (\ref{psidelta}). Let $z$ be small enough so that (\ref{p-forAss4}) holds true.
Then  Assumption~4 is valid with $a(X)= N(X) + |X|,\,\,b(X)
 = \frac{1}{2} \Re  v(X) +\frac{1}{2}\beta ||\psi|| | X |$ and $p=p(z)$ given by \ref{p-forAss4}.
\end{prop}
\textbf{Proof}
Let $X \in {\mathcal X}(B_0(R)),\,\,R>0,$ then for all $r>0$ we have that
\begin{multline}
L(X,z) \equiv \int_{{\mathcal X}^c(B_0(R+r))}\de|\mu_z|(Y) \,  |\zeta(X,Y)| e^{a(Y) + 2b(Y)} a(Y)\\
=\int_{{\mathcal X}^c(B_0(R+r))}\de|\mu_z|(Y) \,
1_{{\mathcal X}(B^c_0(R+\frac{r}{2}))}(Y) |\zeta(X,Y)| e^{a(Y) + 2b(Y)} a(Y) \\
+\int_{{\mathcal X}^c(B_0(R+r))}\de|\mu_z|(Y) \,  1_{{\mathcal X}^c(B^c_0(R+\frac{r}{2}))}(Y)
|\zeta(X,Y)| e^{a(Y) + 2b(Y)} a(Y)\\ \equiv  L_1(X,z) + L_2(X,z).
\end{multline}

We consider first  $L_1(X,z)$. In this case the hard core does not play any role therefore $L_1(X,z)$ can be treated similarly to
the quantity $J''(X,z)$ from (\ref{Ass3J(X,z)}). It follows from  \ref{estmodMayer}  
that for all $ Y^0+s\in {\mathcal X}(B^c_0(R+\frac{r}{2})) $ and all $X\in {\mathcal X}(B_0(R))$,
\begin{equation}
\sum_{s\in B^c_0(R+\frac{r}{2})}|u(X,Y^0+s)| \leq \sum_{x\in X}\sum_{y^0\in Y^0}\int_0^\beta \de t
\sum_{|s|\geq \frac{r}{2}}|\psi(s)|
\leq\beta |X| |Y^0|||\psi_l|| (1+r)^{-l} 
\end{equation}
Hence by Lemma  \ref{Lem1-old-prop1} and formulas (\ref{Rev-a-b-norm}), (\ref{inta(Y^0)})
we find that
\begin{equation}
L_1(X,z)
\leq \frac{1}{2} |X| \beta ||\psi_l|| (1+\beta e M)  
\sum_{j=1}^\infty j \left[q(z)e^{\beta||\psi||}\right]^j (1+r)^{-l}.
\end{equation}

Now we consider $L_2(X,z)$. Here we have to take into account that the interaction has a hard core. The treatment is similar to that
of $J(X,z)$ from \ref{Ass3J(X,z)} but here instead of Lemma \ref{Lem1-old-prop1} we use Lemma \ref{Lemma2} and the inequality
(\ref{InegT}). Then
\begin{multline}
L_2(X,z)
\leq
 \sum_{j=1}^\infty \frac{[z  e^{\beta(M+\Re M_0+ ||\psi||) +1}]^j}{j}
\int P^{j\beta}_{00,+}(\de Y^0) e^{N(Y^0))}1_{\sup |Y^0|\geq\frac{r}{2}}(Y^0)\\
\cdot
a(Y^0)\left[\sum_{s\in T(X,Y^0)}\mid \zeta(X, Y^0+s)\mid
+ \sum_{s\in T^c(X,Y^0)}\mid \zeta(X, Y^0+s)\mid  \right]\\
\equiv L'_2(X,z) + L''_2(X,z).
\label{L_2}
\end{multline}
We start with $L'_2(X,z)$. In view of (\ref{InegT}) it is evident that
\begin{multline}
L'_2(X,z)
\leq  a(X) \sum_{j=1}^\infty \frac{[z  e^{\beta(M+\Re M_0+ ||\psi||) +1}]^j}{j}
\int P^{j\beta}_{00,+}(\de Y^0) e^{N(Y^0))}1_{\sup |Y^0|\geq\frac{r}{2}}(Y^0)
\\
 \cdot a^2(Y^0).
\end{multline}
Since by Lemma \ref{Lemma2}
\begin{multline}
\int P^{j\beta}_{00,+}(\de Y^0) e^{N(Y^0))}1_{\sup |Y^0|\geq\frac{r}{2}}(Y^0) a^2(Y^0)
\leq j^2e^{j\beta(M_l e - M)} \left(1+\frac{r}{2}\right)^{-l}\\
\cdot [(1+\beta M_l e )^2 + \beta M_l e ]
\end{multline}
we find that
\begin{equation}
L'_2(X,z)
\leq a(X)   [(1+\beta M_l e )^2 + \beta M_l e ]
  \sum_{j=1}^\infty j \left[q_l(z)e^{\beta ||\psi||}\right]^j \left(1+\frac{r}{2}\right)^{-l}.
 \label{L'_2}
\end{equation}

To treat $L''_2(X,z)$ we use again Lemma \ref{Lemma2} and inequality (\ref{Lem2-old-prop2})
and get
\begin{equation}
L''_2(X,z)
\leq |X|   \beta ||\psi||(1+ \beta M_l e)
  \sum_{j=1}^\infty j \left[q_l(z)e^{\beta ||\psi||}\right]^j \left(1+\frac{r}{2}\right)^{-l}.
\label{L''_2}
\end{equation}
Now combining (\ref{L_2}),  (\ref{L'_2}) and (\ref{L''_2}) we find that
\begin{equation}
L(X,z)
\leq a(X)  C(\beta,l)
  \sum_{j=1}^\infty j \left[q_l(z)e^{\beta ||\psi||}\right]^j \left(1+\frac{r}{2}\right)^{-l}
\label{L(X,z)}
\end{equation}
where $C(\beta,l)$ is given by (\ref{Const.inAss4}).
This completes the proof of Proposition \ref{prop4-oldprop5}.

\subsection*{Appendix B: decay of correlations}

Proposition \ref{SigmaIntegr} and Lemma \ref{prop2-oldLemma1} are proved in \cite{P2}.
To make the paper self contained we give below the sketch of the proofs of these results.
\begin{prop}
\label{SigmaIntegr}
If Assumption 1 and 3 hold true then for all admissible $X \in \mathcal{X}$
\begin{equation}
\int\de|\mu_z|(Y) \, |\sigma(X,Y)| a(Y) \leq e^{a(X)+2b(X)}  a(X) \frac{p}{1-p}.
\end{equation}
\end{prop}

\textbf{Proof.}
According to Theorem 2.3 from \cite{PU}
if Assumptions 1 and 2 hold true, we have for all admissible $X,Y \in {\mathcal X}$,
\begin{multline}
|\sigma(X,Y)| \leq e^{a(Y)+2b(Y)}  \sum_{m\geq0}
\int_{{\mathcal X}^m}\prod_{i=1}^m|\mu_z|(\de X_i)   \prod_{i=0}^m |\zeta(X_i,X_{i+1})| e^{a(X_i)+2b(X_i)}
\label{twopoint}
\end{multline}
with $X_0 \equiv X$ and $X_{m+1} \equiv Y$. The term of the series corresponding to $m=0$ is $ |\zeta(X,Y)| e^{a(X)+2b(X)}$ by definition.
 Then it follows with the help of Assumption 3 and Monotone Convergence Theorem that
\begin{equation}
\int_{\mathcal X} |\mu_z|(\de Y) |\sigma(X,Y)|
\leq e^{a(X)+2b(X)} a(X) \sum_{m\geq0}p^{m+1}.
\end{equation}
This completes the proof of Proposition \ref{SigmaIntegr}.

\textbf{Proof of Proposition \ref{MainProp1}.} The proof of Proposition \ref{MainProp1} is based on the following important result
\begin{lem} (\cite{P2})
If Assumptions 1, 3 and  4  hold true then for all admissible  $X \in {\mathcal X}(B_0(R))$ and all $R, r >0$,
\begin{align}
\int_{{\mathcal X}^c(B_0(R+r))}\de|\mu_z|(Y) \, |\sigma(X,Y)|  \leq C(l,p) a(X) e^{a(X)+2b(X)} (1+r)^{-l}
\end{align}
where $C(l, p) = 2^l \sum_{m=1}^\infty m^{l+1} p^{m}$.
\label{prop2-oldLemma1}
\end{lem}
 \textbf{Proof of Lemma \ref{prop2-oldLemma1}.}
 First we note that Assumption 3 is stronger than Assumption 2, hence under the conditions of Proposition \ref{MainProp1} we can use
 the formula (\ref{twopoint}). Then
\begin{equation}
\int_{{\mathcal X}^c(B_0(R+r))} |\mu_z|(\de Y) |\sigma(X,Y)| \leq  e^{a(X)+2b(X)}  \sum_{m\geq0} D_m(X, R, r)
\end{equation}
where
\begin{align}
D_m(X, R, r)=&\int_{{\mathcal X}^c(B_0(R+r))} |\mu_z|(\de Y)
 e^{a(Y)+2b(Y)} \int_{{\mathcal X}^m}\prod_{i=1}^m|\mu_z|(\de X_i) e^{a(X_i)+2b(X_i)} \nonumber\\
 \cdot& \prod_{i=0}^m |\zeta(X_i,X_{i+1})|  ,\,\,m\geq1
 \label{Dmdef}
\end{align}
and
\begin{equation}
D_0(x, R, r)=\int_{{\mathcal X}^c(B_0(R+r))}\de|\mu_z|(Y) e^{a(Y)+2b(Y)}  |\zeta(X,Y)|.
\end{equation}

One can prove by induction in $m$ that for any $X\in{\mathcal X}(B_0(R))$ and $r>0$,
\begin{align}
D_m(X, R, r)\leq a(X)(m+1) p ^{m+1}\Bigl(1+\frac{r}{2(m+1)}\Bigr)^{-l}.
\label{Dm}
\end{align}
This completes the proof of Lemma \ref{prop2-oldLemma1}.

Denoting the left hand side of \ref{boundMainProp1} by $K(R,z)$ we can decompose it as
\begin{multline}
K(R,z)
= \sum_{j=1}^\infty \frac{[z  e^{\beta(M+\Re M_0)}]^j}{j} \int  P^{j\beta}_{00,+} (\de X^0) e^{-\Re v(X^0)}
1_{{\mathcal X}(B_0(\frac{R}{3}))}(X^0)
 \int_{{\mathcal X}^c(B_0(R))}\\
 \cdot\de|\mu_z|(Y) \, |\sigma(X^0,Y)|
 + \sum_{j=1}^\infty \frac{[z  e^{\beta(M+\Re M_0)}]^j}{j}
 \int  P^{j\beta}_{00,+} (\de X^0) e^{-\Re v(X^0)}\\
 \cdot 1_{{\mathcal X}^c(B_0(\frac{R}{3}))}(X^0)
 \int_{{\mathcal X}^c(B_0(R))}\de|\mu_z|(Y) \, |\sigma(X^0,Y)|
 \equiv K_1(R,z)  +  K_2(R,z).
\end{multline}
With the help of Proposition \ref{prop2-oldLemma1}, formulae (\ref{Rev-a-b-norm}) and (\ref{inta(Y^0)}) we can write
\begin{multline}
K_1(R,z)
\leq  C(l, p) \left(1+\frac{2R}{3}\right)^{-l}
 \sum_{j=1}^\infty \frac{[z  e^{\beta(M+\Re M_0 + ||\psi||) +1}]^j}{j}
 \int  P^{j\beta}_{00,+} (\de X^0)\\
\cdot e^{N(X^0)} a(X^0)
 \leq
   \left(1+R\right)^{-l} C(l, p) (1+ \beta e M ) \sum_{j=1}^\infty [q(z)]^j
\end{multline}

Next consider $K_2(R,z)$. By (\ref{Rev-a-b-norm}) and  Lemma \ref{Lemma2}  we have
\begin{multline}
 K_2(R,z)\leq \frac{p}{1-p} \sum_{j=1}^\infty \frac{[z  e^{\beta(M+\Re M_0 + ||\psi||) +1}]^j}{j}
 \int  P^{j\beta}_{00,+} (\de X^0) 1_{{\mathcal X}^c(B_0(\frac{R}{3}))}(X^0)\\
e^{N(X^0)}  a(X^0)
\leq (1+R)^{-l}  3^l    \frac{p}{1-p} (\beta e M_l + 1)
\sum_{j=1}^\infty [ q_l(z)]^j
\end{multline}

Thus
\begin{equation}
 K(R,z)\leq (1+R)^{-l}  C(l,p) (1+\beta e M_l)
\sum_{j=1}^\infty [ q_l(z)]^j .
\end{equation}
This completes the proof of
Proposition \ref{MainProp1}.

\bibliographystyle{unsrt}

\end{document}